\documentclass[
 aip,
 preprint,%
]{revtex4-1}
\usepackage{amsmath}
\usepackage{graphicx}
\usepackage{dcolumn}
\usepackage{bm}

\usepackage[utf8]{inputenc}
\usepackage[T1]{fontenc}
\usepackage{mathptmx}
\usepackage{etoolbox}
\usepackage{bm}

\makeatletter
\def\@email#1#2{%
 \endgroup
 \patchcmd{\titleblock@produce}
  {\frontmatter@RRAPformat}
  {\frontmatter@RRAPformat{\produce@RRAP{*#1\href{mailto:#2}{#2}}}\frontmatter@RRAPformat}
  {}{}
}%
\makeatother
\begin{document}

\preprint{AIP/123-QED}

\title[BOUT-MTM]{Theoretical and Global Simulation Analysis of Collisional Microtearing Modes}
\author{K. Fan}
 \affiliation{School of Physics, Peking University, Beijing 100871, China.}
 \affiliation{Institute of Plasma Physics, Chinese Academy of Sciences, Hefei 230031, China.}

\author{X.Q. Xu*}%
 \email{xu2@llnl.gov}
 \affiliation{Lawrence Livermore National Laboratory, Livermore, CA 94550 USA.}

\author{B. Zhu}
\affiliation{Lawrence Livermore National Laboratory, Livermore, CA 94550 USA.}%

\author{C. Dong}
\affiliation{Institute of Physics, Chinese Academy of Sciences, Beijing 100190 China.}%

\author{T. Xia}
\affiliation{Institute of Plasma Physics, Chinese Academy of Sciences, Hefei 230031, China.}%


\author{Z. Li}
\affiliation{General Atomics, San Diego, CA 92121, USA.}%



\date{\today}

\begin{abstract}

This study delves into Microtearing Modes (MTMs) in tokamak plasmas, employing advanced simulations within the BOUT++ framework. The research, centering on collisional MTMs influenced by the time-dependent thermal force, enhances our understanding of plasma dynamics. It achieves this through the simplification and linearization of control equations in detailed linear simulations. The study meticulously evaluates various conductivity models, including those proposed by Larakers, Drake, and Hassam, under diverse plasma conditions and collision regimes. A notable achievement of this research is the derivation of a unified dispersion relation that encompasses both MTM and Drift-Alfven Wave (DAW) instabilities. It interestingly reveals that DAW and MTM exhibit instability at different proximities to the rational surface. Specifically, MTMs become unstable near the rational surface but stabilize farther away, whereas the drift-Alfven instability manifests away from the rational surface. Further, the study re-derives MTM dispersion relations based on Ohm's law and the vorticity equation, providing a thorough analysis of electromagnetic and electrostatic interactions in tokamaks. Global simulations demonstrate an inverse correlation between MTM growth rates and collisionality, and a direct correlation with temperature gradients. The nonalignment of the rational surface with the peak $\omega_{*e}$ stabilizes the MTMs. Nonlinear simulations highlight electron temperature relaxation as the primary saturation mechanism for MTMs, with magnetic flutter identified as the dominant mode of electron thermal transport.

\end{abstract}

\maketitle

\section{Introduction}\label{sec:intro}

The microtearing mode (MTM) is an electromagnetic micro-instability that significantly affects thermal transport and energy confinement in the tokamak plasma. In standard tokamak H-mode discharges, such as those observed in DIII-D~\cite{nelson2021time}, JET~\cite{hatch2021microtearing}, and others, MTMs are frequently observed. MTM can dominate electron thermal transport in the JET ITER-like wall pedestal under certain conditions~\cite{hatch2016microtearing}, plays a role in determining the behavior of electron temperature during an ELM cycle~\cite{chen2023micro}, and in double barrier operations, it generates a substantial electron thermal transport at the Inner Transport Barrier (ITB)\cite{jian2019role}. In the high-$\beta$ regime, MTMs may become more prominent as KBMs enter their second stable regime, thereby playing a significant role alongside the behavior of KBMs. Therefore, MTMs can become the dominant low-$k_\theta$ modes as $\beta$ increase in spherical tokamak devices like NSTX\cite{kaye2014reduced,guttenfelder2011electromagnetic}.

MTM was initially recognized as a temperature-gradient-driven tearing mode in the ST Tokamak in the 1970s. Using a variant method, Hazeltine et al. ~\cite{hazeltine1975kinetic}first proposed a linear kinetic conductivity model for this type of mode. This kinetic theory was further developed by Drake et al.~\cite{drake1977kinetic}, with the model being extended to a semi-collisional regime. More recently, Larakers et al.\cite{larakers2020comprehensive,larakers2021global} have successfully expanded the theory to include the global radial variation effect, providing a comprehensive explanation for the separate bands observed in the MTM frequency spectrum.



BOUT++ is a C++ framework specifically developed for solving initial value partial differential equations\cite{dudson2009bout++} related to magnetized plasmas. It ensures a clear delineation between numerical and physics development coding, thereby enhancing the efficiency of the simulation code development process. Successfully applied in tokamak boundary plasma simulations~\cite{xi2014phase, xu2019simulations}, including edge-localized modes, boundary turbulence, and blobs. Plasma models in BOUT++ have been validated with experimental data and results from other codes.

In this work, we have extended the capabilities of the BOUT++ code to include the simulation of the MTM. This paper presents our global simulation of the collisional MTM in a shifted-circle tokamak geometry. We have implemented Hassam's fluid model for the MTM~\cite{hassam1980higher}, incorporating electron inertia and a time-dependent thermal force into Ohm's law equation. Both global linear and nonlinear simulations are performed to investigate the MTM characteristics. 


This paper is organized as follows. In Section \ref{sec:theory}, a brief survey of existing theoretical models on MTM is provided. Section \ref{sec:model and simu setup} describes the simulation model and parameters used in this study in detail. Linear simulation results are summarized in Section \ref{sec:global linear simu} as we explore the mode structures and the relation between frequency and growth rate spectrum and collisionality and temperature gradient. The effect of alignment on the growth rate is investigated as well. Nonlinear MTM simulations are presented in Section \ref{sec:nonlinear global simu}. We find the saturation mechanism of MTMs due to temperature profile relaxation and compare the heat transport resulting from $E\times B$ convection and magnetic flutter. Finally, Section \ref{sec:summary} summarizes our findings.

\section{Theory of microtearing mode}\label{sec:theory}
In this section, we delve into the theory of MTMs, examining its dispersion relation and exploring its connection to classical tearing modes as well as DAW instabilities.

\subsection{\label{sec:review}Review and comparison of MTM models}
MTM is related to the resistive current layer physics. Different models are characterized by the conductivity, $\sigma$, in Ohm's law.
\begin{equation}
    j_{\parallel} = \sigma E_{\parallel}
\end{equation}
Recently, the classical Hazeltine electric conductivity model~\cite{hazeltine1975kinetic} has been extended to include radial variation by Larakers et.al \cite{larakers2021global}. If using a Lorentz Gas collision operator neglecting the electron-electron collision,
\begin{equation}
    C_{LG}(f)=\frac{\nu_{ei}}{2v^3} \frac{\partial}{\partial \xi} (1-\xi^2) \frac{\partial}{\partial \xi} f 
\end{equation}
where $\xi=v_{\parallel}/ v$ is the pitch angle, $ \nu_{ei}$ is the collision rate, $\nu_{ei}=16\sqrt{\pi} n_e Z e^4 \text{ln} \Lambda /3 m_e^{1/2}v_{te}^{3}$, where $v_{te}$ is the electron thermal speed, the conductivity has a form of,
\begin{equation}
    \sigma = -\frac{2e^2n_e}{\nu_{ei}m_e}\left[(1 - \frac{\omega_{*ne}}{\omega}-\frac{\omega_{*Te}}{\omega})L_{11} - \frac{\omega_{*Te}}{\omega}L_{12}\right],
    \label{eq: sigma Larakers}
\end{equation}
\begin{equation}
    L_{11} = \frac{4}{3\sqrt{\pi}} \int_0^\infty \frac{s^4e^{-s^2}ds}{i\hat{\omega}-\frac{i\hat{k}^2s^2}{3\hat{\omega}} -\frac{1}{s^3} + \frac{4}{15} \frac{\hat{k}^2s^2}{\alpha_n}},
\end{equation}
\begin{equation}
    L_{12} = \frac{4}{3\sqrt{\pi}} \int_0^\infty \frac{s^4(s^2-5/2)e^{-s^2}ds}{i\hat{\omega}-\frac{i\hat{k}^2s^2}{3\hat{\omega}} -\frac{1}{s^3} + \frac{4}{15} \frac{\hat{k}^2s^2}{\alpha_N}}.
\end{equation}
\begin{equation}
    \begin{aligned}
        \alpha_N = i\hat{\omega} - \frac{N(N+1)}{2s^3} + \frac{(N+1)^2}{(2N+1)(2N+3)} \frac{\hat{k}^2s^2}{\alpha_{N+1}}
    \end{aligned}
\end{equation}

and $\hat{\omega}=\omega/\nu_{ei}$, $\hat{k}=k_{\parallel}v_{te}/\nu_{ei}$, $\omega_{*ne}= \frac{k_\theta T_e}{eB}L_n^{-1}$, $ \omega_{*Te}= \frac{k_\theta T_e}{eB}L_T^{-1}$, where $L_n=|\nabla n_e|/n_e,L_T=|\nabla T_e|/T_e$ are characteristic lengths of electron density and temperature gradients, $k_\theta$ is the poloidal wave number. In the fluid response limit, $k_{\parallel}v_{te}\ll \omega,\nu_{ei}$, the coefficients are,  
\begin{equation}
    L_{11} =  \frac{4}{3\sqrt{\pi}} \int_0^\infty \frac{s^4e^{-s^2}ds}{i\hat{\omega}-\frac{1}{s^3}  }, 
\end{equation}

\begin{equation}
    L_{12} =  \frac{4}{3\sqrt{\pi}} \int_0^\infty \frac{s^4(s^2-5/2)e^{-s^2}ds}{i\hat{\omega}-\frac{1}{s^3}  }. 
\end{equation}
Then Larakers' model in different limits is compared with Drake\cite{drake1980microtearing} and Hassam's \cite{hassam1980fluid}models.

In the $\omega\ll\nu_{ei}$ limit, the coefficients $L_{11},L_{12}$ reduce to,
\begin{equation}
    L_{11} = -\frac{4}{3\sqrt{\pi}}\left(3+\frac{945\sqrt{\pi}}{64}i\frac{\omega}{\nu_{ei}}  \right) ,
\end{equation}
\begin{equation}
    L_{12} = -\frac{4}{3\sqrt{\pi}}\left(\frac{9}{2}+\frac{2835\sqrt{\pi}}{64}i\frac{\omega}{\nu_{ei}}      \right).
\end{equation}
After substituting into Eq.(\ref{eq: sigma Larakers}), the conductivity is,
\begin{equation}
\begin{aligned}
     \sigma = \frac{32e^2 n_e}{3 m_e \pi \nu_{ei}}\left[\left(1 - \frac{\omega_{*ne}}{\omega}-\frac{\omega_{*Te}}{\omega}\right)\left(1+i\frac{105}{16} \frac{\omega}{\nu_{ei}}\right) - \frac{\omega_{*Te}}{\omega} \left(\frac{3}{2}+i\frac{315}{16} \frac{\omega}{\nu_{ei}}\right) \right], 
\end{aligned}
\end{equation}
which has the same form as Drake's model and Hassam's model.

In the $\omega\gg\nu_{ei}$ limit, following the similar way, the coefficients are
\begin{equation}
    L_{11} = \frac{4}{3\sqrt{\pi}} \left(-i \frac{3\sqrt{\pi}}{8} \frac{\nu_{ei}}{\omega}- \frac{\nu_{ei}^2}{2\omega^2}  \right),
\end{equation}
\begin{equation}
    L_{12} = \frac{1}{\sqrt{\pi}} \frac{\nu_{ei}^2}{\omega^2},
\end{equation}
After substituting into Eq.(\ref{eq: sigma Larakers}), the conductivity is,
\begin{equation}
    \begin{aligned}
       \sigma = -i\frac{e^2 n_e}{m_e \nu_{ei}} \left[ \left(1 - \frac{\omega_{*ne}}{\omega}-\frac{\omega_{*Te}}{\omega}\right)\left(\frac{\nu_{ei}}{\omega}-i\frac{{\nu_{ei}}^2}{\omega^2}\right)
       -i\frac{\omega_{*Te}}{\omega}\frac{3{\nu_{ei}}^2}{2\omega^2} \right]. 
\end{aligned}
\end{equation}
It is the same as Drake's conductivity in the low collisional regime.

In summary, for local assumption around the rational surface, Larakers' conductivity model reduces to Drake's model in both $\omega\ll\nu_{ei}$ and $\omega \gg \nu_{ei}$ limits, and reduces to  Hassam's model in $\omega\ll\nu_{ei}$ limit. Larakers' model thus bridges the two limits and also includes $k_{\parallel}$ effects.

\begin{figure*}
    \centering
    \includegraphics[width=0.6\textwidth,keepaspectratio]{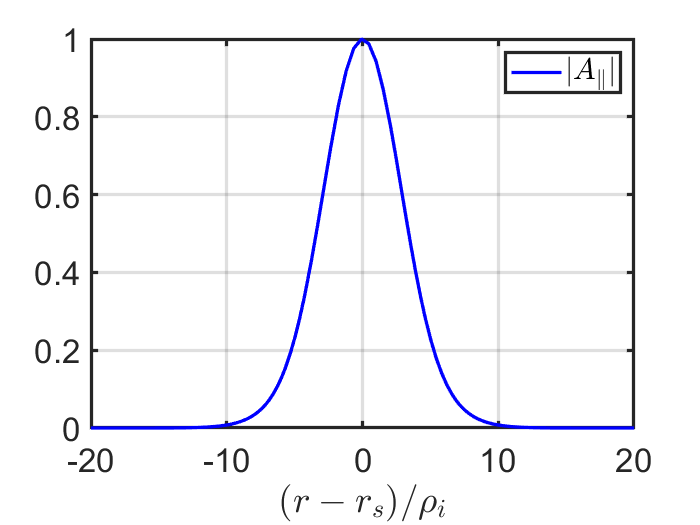}
    \caption{Diagram of the normalized distribution of absolute parallel magnetic vector potential, $A_\parallel$, calculated from Eq.(\ref{eq: first-order dispersion relation}) using a shooting method. The horizontal axis denotes the distance with respect to the rational surface. The parameters are, $\beta=1\% $, $\omega_{*e}/\nu_{ei}=0.18e^{\frac{x^2}{\delta^2}}$ , $\delta=10$, $k_y=0$.}
    \label{fig: nonlocal clc Apar}
\end{figure*}

\subsection{\label{sec:dispersion MTM}Dispersion relations of MTM}

The dispersion relation of MTM is derived from Ohm's law and vorticity equation (or the quasi-neutral condition) and has already been performed under slab or other geometries by many researchers~\cite{hazeltine1975kinetic,drake1977kinetic,hassam1980fluid}. 
If performed in a slab geometry, after linearization and Fourier transformation, the vorticity equation and Ohm's law take the following forms,
\begin{equation}
    \begin{aligned}
        \left( \frac{\partial^2}{\partial x^2}-k_y^2\right) \widetilde{A_\parallel}\  = -\mu_0\sigma(i\omega \widetilde{A_\parallel}-ik_\parallel \widetilde{\phi}),
        \label{eq: slab Ohm's law}
    \end{aligned}
\end{equation}
\begin{equation}
    (\omega-\omega_{*i})\left( \frac{\partial^2}{\partial x^2}-k_y^2\right)\widetilde{\phi} = -k_\parallel V_A^2\mu_0 \sigma(i\omega \widetilde{A_\parallel}-ik_\parallel\widetilde{\phi}),
    \label{eq: slab vorticity eq}
\end{equation}
where the tilde means perturbed value, $x$ denotes the radial distance from the rational surface, and $y$ is the orthogonal direction on the magnetic surface. $V_A=B/\sqrt{\mu_0 m_i n_i}$ is the Alfvén velocity, $\mu_0$ is the vacuum permeability. $k_\parallel=k_y x/L_s$ is the parallel wave number, $k_y$ is the binomial wave number, approximately $k_y=m/r$, and $m$ is the poloidal mode number, $L_s$ is the local magnetic shear length. Using a full collision operator, the conductivity of Hassam's mode is,
\begin{equation}
    \sigma = \frac{e^2n_e}{0.51\nu_{ei}m_e\omega}\left[\omega - \omega_{*ne}-(1+\alpha)\omega_{*Te}-i\alpha \alpha^\prime \omega_{*Te} \frac{\omega}{\nu_{ei}}\right],
    \label{eq: conductivity Hassam expression}
\end{equation}
where, $\alpha=0.8, \alpha\prime=0.54$ are constant numbers. These coupled equations can be solved in the inner and outer regions separately, and using a matching condition to obtain the dispersion relation. 

In the inner region, the electromagnetic perturbation dominates. The electrostatic potential plays a role in short-circuiting the electric field. When $E_\parallel \equiv i\omega \widetilde{A_\parallel}- ik_\parallel\widetilde{\phi}=0$, a characteristic length can be solved, $\Delta_E=\omega \widetilde{A_\parallel}L_s/k_y\widetilde{\phi}$, where $L_s$ is the magnetic shear length. Out of the region, $|x|>\Delta_E$, the electric field is short-circuited. If $\Delta_E \gg \Delta_j$, where $\Delta_j$ is the width of the current layer determined by $\sigma(x)$, Eqs. (\ref{eq: slab Ohm's law})  reduces to the equation,
\begin{equation}
    \left( \frac{\partial^2}{\partial x^2}-k_y^2\right)A_\parallel\  = -i\mu_0\sigma\omega A_\parallel.
    \label{eq: first-order dispersion relation}
\end{equation}
The validation of this assumption is $\left(\frac{\omega_{*Te} \nu_{ei}}{\omega_{*ne}^2} \right) \left(\frac{m_e L_s^2}{m_i L_n^2} \right)\ll 1$ \cite{gladd1980electron}. Using a shooting method, the mode structure is obtained and shown in FIG.\ref{fig: nonlocal clc Apar}. 

However, across the current layer, there is a discontinuity of $A_\parallel^\prime$ due to the current sheet, and the prime denotes derivative in the radial direction. This discontinuity can be represented by a tearing parameter, calculated under the constant-$A_\parallel$ assumption, $\Delta^\prime=[A_\parallel^\prime(0+)-A_\parallel^\prime(0-)]/A_\parallel(0)$. The $\Delta^\prime$ is solved in the outer region. Approximately $\widetilde{A_\parallel} \sim e^{-k_y|x|}$, and $\Delta^\prime=-2k_y$\cite{d1980linear}. Then the solution of Eq.(\ref{eq: first-order dispersion relation}) can be obtained from an integration within the current layer width, $\Delta_j$\cite{gladd1980electron}.

Furthermore, in general cases, without $\Delta_E \gg \Delta_j$ assumption and considering the current drive, $J^\prime_{0\parallel}$,  of that classical tearing mode, it is necessary to combine with Eq.(\ref{eq: slab vorticity eq}) to obtain the dispersion relation in a form, 
\begin{equation}
    \begin{aligned}
        \omega^2 \left[\omega - \omega_{*ne}-(1+\alpha)\omega_{*Te}-i\alpha \alpha^\prime \omega_{*Te}\frac{\omega}{\nu_{ei}}\right]^3=i \gamma_c^5,
        \label{eq: Hassam dispersion relation with Delta prime}
    \end{aligned}
\end{equation}
where $\gamma_c=\left[ \frac{\Gamma(1/4)\Delta^\prime L_s}{\pi \Gamma{(3/4)}} \right]^{4/5}\tau_R^{-3/5}\tau_A^{-2/5}$ is the growth rate of classical collisional tearing mode\cite{furth1963finite}. We do not repeat the derivation but just show the result here. This shows that MTM could be unstable even when $\Delta^\prime$ is negative.

When $\omega_{*e}\gg \gamma_c$, the influence of $\gamma_c$ on the solution of Eq.(\ref{eq: Hassam dispersion relation with Delta prime}) is only a minor correction to its main value. It is applicable under most present tokamak operation parameters. And the main value of the solution can be solved from,\cite{hazeltine1975kinetic, larakers2021global}
\begin{equation}
    \sigma = 0,
\end{equation}
This indicates that a dispersion relation $\sigma=0$ can be used to calculate the mode frequency and growth rate.  

\begin{figure*}
    \centering
    \includegraphics[width=1.0\textwidth,keepaspectratio]{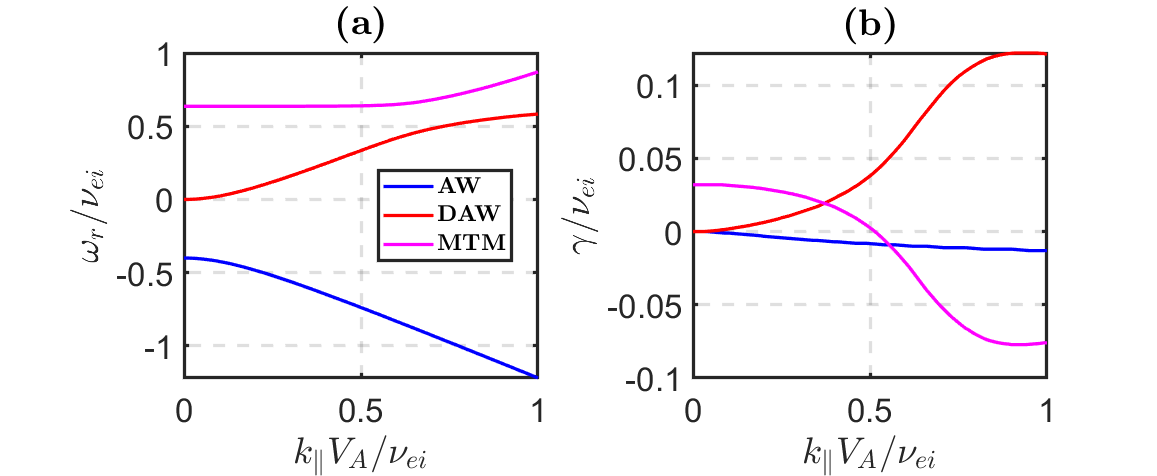}
   \caption{Numerical solution of dispersion relation Eq.(\ref{eq: local dispersion of MTM with kperp}). Panels (a) and (b) show the frequency and growth rate versus normalized $k_\parallel$.}
    \label{fig-DAWvsMTM}
\end{figure*}

If performed in local analysis, it becomes evident that the mode transitions from a microtearing mode to a drift Alfvén wave instability with an increase in $k_\parallel$. Let $\nabla_\bot\longrightarrow-ik_\bot$, Eqs. (\ref{eq: slab Ohm's law})-(\ref{eq: slab vorticity eq}) are reduced to a dispersion relation,
\begin{equation}
    \begin{aligned}
        -i\frac{\eta}{\mu_0}k_\bot^2 \omega & \left(\omega-\omega_{*i} \right)=\left[\omega-\omega_{*ne}-(1+\alpha)\omega_{*Te} -i\frac{\alpha\alpha^\prime\omega}{\nu_{ei}}\omega_{*Te}  \right] \\
        & \times \left[\omega(\omega-\omega_{*i})-k^2_{\parallel}V_A^2 \right].
        \label{eq: local dispersion of MTM with kperp}
    \end{aligned}
\end{equation}
Choosing a set of parameters, $k_\bot^2 \eta/\mu_0 \nu_{ei} = 0.05,\ \omega_{*Te}/\nu_{ei}=3\omega_{*ne}/\nu_{ei}=0.3$, the dispersion relation Eq.(\ref{eq: local dispersion of MTM with kperp}) can be solved numerically in slab geometry. As shown in FIG.\ref{fig-DAWvsMTM}, there are 3 modes. The red curve is recognized as drift Alfvén wave instability. It is unstable with finite $k_\parallel$, in other words, this instability arises at the location where is a little bit shifted away from the rational surface. As shown in FIG.\ref{fig-DAWvsMTM}(b), there is another mode becoming unstable in the small $k_\parallel$ regimes and becoming stabilized further away from the rational surface. This instability is recognized as an MTM driven by the time-dependent thermal force. DAW is unstable with finite $k_\parallel$, as a result, DAW and MTM are unstable at different locations to the rational surface. 

\section{Fluid model and simulation setup}\label{sec:model and simu setup}
In this section, we introduce the control equations, equilibrium profiles, and parameters used in this study.

\subsection{\label{sec:simulation model} Microtearing mode model for simulation}
To minimize the impact of existing fluid-based turbulence model~\cite{zhu2021drift} within BOUT++ framework, Hassam's nonlinear fluid model~\cite{hassam1980higher} is adapted for the global MTM simulation. In this model, the electron parallel momentum equation includes a time-dependent thermal force $\partial(\nabla_\parallel T_e)/\partial t$ term and thus is slightly different from the original Braginskii model~\cite{braginskii1965transport}, 
\begin{equation}
    \begin{aligned}
    \alpha^{\prime\prime} \frac{\partial}{\partial t} \left( m_e n_e V_{e\parallel} \right) = en_e \left(\nabla_\parallel \widetilde{\phi} + \frac{\partial}{\partial t} A_\parallel  \right) - \nabla_\parallel p_e - \alpha n_e \nabla_\parallel T_e + en_e\eta j_\parallel + \alpha\alpha^{\prime} \frac{n_e}{\nu_{ei}} \frac{\partial}{\partial t}\nabla_\parallel T_e
        \label{eq: electron momentum eq of Hassam}
    \end{aligned}
\end{equation}
where $\eta$ is the parallel resistivity. Replacing $A_\parallel$ with $B\widetilde{\psi}$ and dividing $en_eB$ for both sides of Eq.(\ref{eq: electron momentum eq of Hassam}), this equation becomes,
\begin{equation}
    \begin{aligned}
        \frac{\partial}{\partial t} \left( \widetilde{\psi} -  \frac{\alpha^{\prime\prime}m_e}{e B}V_{e\parallel} + \frac{ \alpha\alpha^{\prime}}{ e\nu_{ei}B}\nabla_{\parallel}T_e \right) = -\frac{1}{B}\nabla_{\parallel}\widetilde{\phi} - \frac{\eta}{B}j_{\parallel} +  \frac{1}{en_e B}\nabla_{\parallel}p_e  +  \frac{\alpha}{eB}\nabla_{\parallel}T_e.
        \label{eq: reduced electron momentum eq of Hassam}
    \end{aligned}
\end{equation}
Here we take the semi-electromagnetic assumption~\cite{zhu2023electromagnetic} and the parallel gradient is written as $\nabla_\parallel = \nabla_{\parallel 0} - \textbf{\textit{b}}\times \nabla\widetilde{\psi}$.

Because the MTM is driven by temperature gradients, we simplify the model to include the fundamental physics and combine it with the vorticity equation in \textit{6-field} model\cite{zhu2021drift} that with finite ion Larmor radius effect (denoted by the gyro-viscosity terms) included. Finally, the control equations are reduced to a three-field model,
\begin{equation}
\begin{aligned}
      \frac{\partial}{\partial t} U =\  & -\bm{V}_E \cdot \nabla U + B^2 \nabla_{\parallel}\left(\frac{J_{\parallel}}{B}\right) + 2\bm{b}\times\bm{\kappa}\cdot\nabla P 
      + \mu_{\parallel}\nabla^2_{\parallel}U + \mu_{\bot}\nabla^2_\bot U \\
      & -\frac{1}{2\Omega_i} \left[ \frac{1}{B } \bm{b}\times \nabla P_i \cdot \nabla (\nabla_\bot^2 \Phi) - Z_i eB \bm{b}\times \nabla n_i \cdot \nabla \left( \frac{\nabla \Phi}{B} \right)^2  \right] \\
      &+ \frac{1}{2\Omega_i} \left[ \frac{1}{B} \bm{b}\times\nabla\Phi \cdot \nabla(\nabla^2_\bot P_i) - \nabla_\bot^2 \left( \frac{1}{B}\bm{b}\times\nabla\Phi\cdot\nabla P_i   \right) \right],
      \label{eq: vorticity evolution}
\end{aligned}
\end{equation}
\begin{equation}
\begin{aligned}
    \frac{\partial}{\partial t} A_{j \parallel} = -\frac{1}{B}\nabla_{\parallel}\Phi -C_{\nu}\frac{\eta}{B}j_{\parallel} + C_T\frac{1}{en_e B}\nabla_{\parallel}p_e  + C_T\frac{\alpha}{eB}\nabla_{\parallel}T_e + \eta_H\nabla^2_\bot j_{\parallel},
    \label{eq: electron momentum}
\end{aligned}
\end{equation}
\begin{equation}
    \begin{aligned}
            \frac{\partial}{\partial t} T_e = -\bm{V}_E \cdot \nabla T_e - V_{e\parallel}\nabla_{\parallel}T_e -\frac{2}{3n_e} \nabla_{\parallel}q_e  -\frac{2}{3}T_e \left[\left(\frac{2}{B}\bm{b}\times\bm{\kappa} \right)\cdot \left(\nabla\Phi - \frac{1}{en_e}\nabla P_e 
            \right)  \right],
            \label{eq: temperature evolution}
    \end{aligned}
\end{equation}
\begin{equation}
    U = \frac{n_i m_i}{B} \left(\nabla^2_{\bot} \widetilde{\phi} + \frac{1}{n_{i} Z_i e}\nabla^2_{\bot}p_i \right) ,
    \label{eq: vorticity expression in full eqs }
\end{equation}
\begin{equation}
   A_{j \parallel} = \widetilde{\psi} - \frac{\alpha^{\prime\prime}m_e}{e B}V_{e\parallel} + \frac{C_T\alpha\alpha^{\prime}}{C_{\nu}e\nu_{ei}B}\nabla_{\parallel}T_e ,
   \label{eq: Ajpar expression}
\end{equation}
\begin{equation}
 J_\parallel = J_{0\parallel} + j_{\parallel},\ 
    j_{\parallel} = -\frac{B_0}{\mu_0}\nabla^2_\bot \widetilde{\psi},
    \label{eq:jpar expression}
\end{equation}

Here, $\textbf{\textit{b}}=\textit{\textbf{B}}/B$ denotes the direction of magnetic field, $\bm{V}_E=\textbf{\textit{b}}\times \nabla\Phi/B$ is the drift velocity, $\Phi=\widetilde{\phi}+\Phi_0 $ is the total electrostatic potential, $\widetilde{\phi}$ and $\Phi_0$ are the perturbed and equilibrium electrostatic potential respectively, $\bm{\kappa}=\textbf{\textit{b}}\cdot\nabla\textbf{\textit{b}}$ is the magnetic curvature, $\Omega_i$ is the ion gyro-frequency, $\mu_{\parallel},\mu_{\bot}$ are the parallel and orthogonal viscosity for numerical stability~\cite{xu2011nonlinear}. Spitzer-Harm resistivity~\cite{braginskii1965transport} is implemented. 
\begin{equation}
    \eta = 0.51 \frac{m_e \nu_{ei}}{e^2 n_e},
\end{equation}
$\eta_H$ is the hyper-resistivity \cite{xu2010nonlinear}. Other coefficients are set as $\alpha=0.8,\alpha^{\prime}=0.54,\alpha^{\prime\prime}=1.24$\cite{hassam1980fluid}. $C_{\nu},C_T $ are two free coefficients used for collisionality and gradient dependence study. In the following section, a scan of these coefficients is performed to test the frequency and growth rate spectrum versus collisionality and temperature gradient. Ion parallel velocity is neglected, therefore, the perturbed current 
is approximately $j_{\parallel}=-en_eV_{e\parallel}$. A flux limit closure\cite{zhu2021drift} with the harmonic average of free streaming and conductivity model for heat flux is used,
\begin{equation}
    q_{e\parallel}=-\kappa_{fl}\nabla_{\parallel}T_e,\ \ \ \kappa_{fl}=\frac{\alpha_H \kappa_{FS}\kappa_{SH}}{\alpha_H \kappa_{FS}+\kappa_{SH}},
    \label{heatflux}
\end{equation}
\begin{equation}
    \kappa_{FS}=n_e v_{te} L_\parallel,\ \ \kappa_{SH} = 3.2\frac{n_e  T_e}{\nu_{ei }m_e}.
\end{equation}
where $\alpha_H$ is a weight factor and we call it \textit{free streaming parameter}, $\kappa_{FS}$ is the free streaming thermal transport coefficient, and $\kappa_{SH}$ is the conductive thermal transport coefficient,\cite{zhu2023electromagnetic} $L_\parallel =q \Bar{L}$ is the parallel correlation length. When $\alpha_H$ is close to infinity, the flux limit model will reduce to the conductivity model. In the simulation performed in this article, this factor is chosen to be $\alpha_H=0.05$ by default. 

If the diamagnetic drift effect is considered (i.e., finite $\bm{\kappa}$), the second term in the right-hand side of vorticity Eq.(\ref{eq: vorticity evolution}) must be included to maintain equilibrium~\cite{zhu2018up}. Note in our simulations, cold ion model was adapted so that the perturbation of ion density and temperature vanish, making the second term in the right-hand side of Eq.(\ref{eq: vorticity expression in full eqs }) always zero. At the same time, the  $E_r$ shear flow, $\mathbf{V}_{E 0}=\frac{\mathbf{b_0}\times\nabla\Phi_0}{B_0} $, is included to balance equilibrium ion diamagnetic drift for particle conservation~\cite{zhu2017global}. Also $E_r$ shear flow introduces the Doppler shift effect, $\frac{\partial}{\partial t} \xrightarrow{} \frac{\partial}{\partial t} + \mathbf{V}_{E0}\cdot\nabla$. Consequently, when referring to the equilibrium ion diamagnetic drift effect, we consider the equilibrium flow induced by equilibrium ion diamagnetic drift, balanced by the $E_r$ shear flow—essentially investigating the impact of $E_r$ shear flow.

\begin{figure*}
    \centering   
    \includegraphics[width=1.0\textwidth,keepaspectratio]{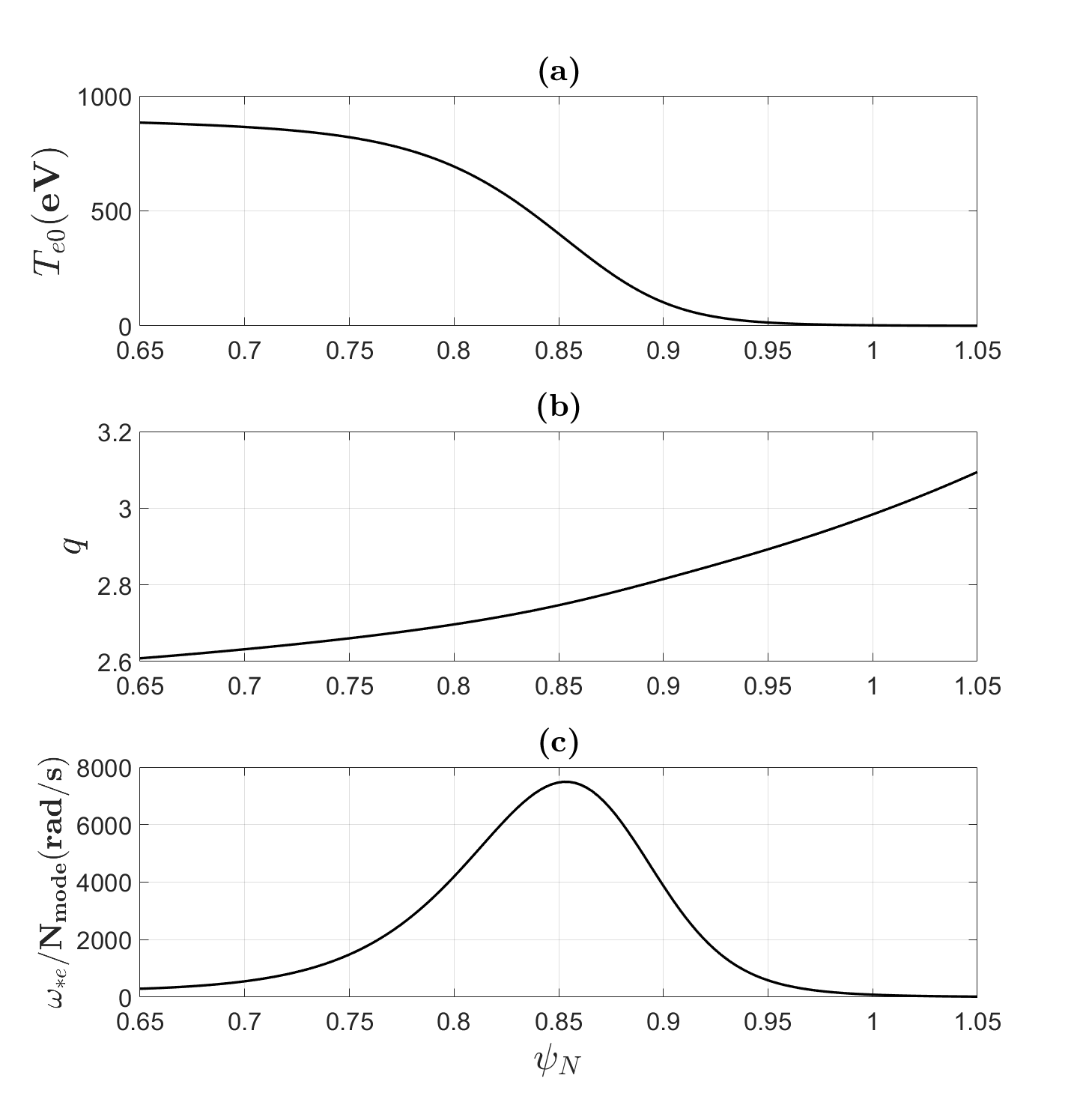}
    \caption{ (a)Electron temperature equilibrium profile, (b)safety factor profile, and (c)$\omega_{*Te}$ profile normalized to its toroidal mode number.}
    \label{fig:Eqprofile}
\end{figure*}

\subsection{Simulation setup}\label{sec:setup} 
In our simulation, we utilize a shifted circular geometry equilibrium generated by CORSICA~\cite{crotinger1997corsica}. The equilibrium profile is displayed in FIG. \ref{fig:Eqprofile}. Our simulation domain extends from the flux surface $\psi_N=0.65$ to $\psi_N=1.05$. To simplify the physics and focus on the impact of temperature gradients, we maintain a constant density of $10^{20} [\mathrm{m^{-3}}]$ throughout the entire domain, thereby avoiding the influence of density gradients.

The safety factor profile with $2.6<q<3.1$ for the simulation domain is intentionally set to be flatter than usual to distinguish rational surfaces clearly for discussions. In contrast to the standard CORSICA-generated equilibrium, our equilibrium features a lower current profile and a stronger magnetic field. It's worth mentioning that Seto et al.~\cite{seto2023bout++} have made significant developments in BOUT++ to enable accurate low-n mode simulations, although this version may come with slower code execution speed. So, we do not perform very low-$n$ modes simulation.

A field-aligned coordinate is implemented in BOUT++ \cite{xu2008boundary}, where $x=\Psi-\Psi_0,y=\theta,z=\zeta-\int_{\theta_0}^\theta \nu (\Psi,\theta)d\theta$, with $\nu$ being the local filed-line pitch. Therefore, $\mathbf{e_x}$ is along the cross flux direction, and $\mathbf{e_y},\mathbf{e_z}$ are along the parallel and bi-normal directions. For linear simulation, the grid points are $\text{NX}=512, \text{NY}=128, \text{NZ}=16$ in $x$, $y$, and $z$ directions respectively, and for nonlinear simulation, $\text{NZ}=32$. It is a twist-shift boundary condition in the $y$ direction and periodic in the $Z$ direction. For $T_e, \ A_{j \parallel}$, the boundary condition is the Neumann condition in the inner boundary and the Dirichlet in the outer boundary; for $U$, it is Dirichlet in both boundaries. Other parameters are: Alfvén time $\tau_A = 8.035\times 10^{-7}[\text{s}]$, normalization length $\overline{L}=3.1574[\text{m}]$, Alfvén speed $V_A=3.9295\times 10^{6}[\text{m/s}]$, normalization magnetic field $\Bar{B}=2.55[\text{T}]$, normalization electron temperature $\Bar{T}_e = 840[\text{eV}]$. 

\section{Global linear simulation of MTM}\label{sec:global linear simu}
For the linear simulation, in order to easily understand the physics, Eqs.(\ref{eq: vorticity evolution})-(\ref{eq:jpar expression}) are further reduced. After linearization, the control equations for linear simulation are,
\begin{equation}
    \begin{aligned}
     \frac{\partial}{\partial t}\widetilde{U}= -V_{E0}\cdot\nabla \widetilde{U} -B_0 \nabla_{\parallel 0} j_{\parallel}-B_0\times \nabla\widetilde{\psi}\cdot\nabla J_{\parallel 0}+2\bm{b_0}\times\kappa\cdot\nabla \widetilde{p},
     \label{eq: linearized vorticity eq}
    \end{aligned}
\end{equation}
\begin{equation}
    \begin{aligned}
     \frac{\partial}{\partial t}\widetilde{A_{j \parallel}} = -V_{E0}\cdot\nabla \widetilde{A_{j\parallel}} -\frac{1}{B_0}\nabla_{\parallel0}\widetilde{\phi}-C_{\nu}\frac{\eta}{B_0}j_{\parallel} +C_T\left(1+\alpha \right) \frac{1}{e B_0}\left(\nabla_{\parallel 0}\widetilde{T_e} -\bm{b}_0\times\nabla\widetilde{\psi}\cdot\nabla T_{e0}\right),
    \end{aligned}
\end{equation}
\begin{equation}
    \begin{aligned}
     \frac{\partial}{\partial t}\widetilde{T_e} = &-V_{E0}\cdot\nabla \widetilde{T}_e-\frac{\bm{b_0} \times\nabla\widetilde{\phi}}{B_0}\cdot\nabla T_{e0}+\frac{2}{3}\nabla_{\parallel 0} \widetilde{T_e}\left(\kappa_{fl}\nabla_{\parallel 0} \widetilde{T_e} -\kappa_{fl}\bm{b}_0\times\nabla\widetilde{\psi}\cdot\nabla T_{e0} \right)\\
    \end{aligned}
\end{equation}
where $\bm{b_0}$ denotes the direction of the unperturbed magnetic field. And Eqs.(\ref{eq: vorticity expression in full eqs }) and (\ref{eq: Ajpar expression}) can be written in forms,
\begin{equation}
    \begin{aligned}
    \nabla_\bot^2\widetilde{\phi}=\frac{B_0}{n_{i0}m_i}\widetilde{U}-\frac{1}{n_{i0}Z_e e}\nabla_\bot^2\widetilde{p_i},
    \label{eq: invert lapalce phi}
    \end{aligned}
\end{equation}
\begin{equation}
    \begin{aligned}
    \frac{\alpha^{\prime\prime}m_e}{e^2n_{e0}\mu_0}\nabla_\bot^2\widetilde{\psi}- \ \widetilde{\psi}+\frac{C_T \alpha\alpha^\prime}{C_\nu e\nu_{ei}B_0}\bm{b}_0\times\nabla\widetilde{\psi}\cdot\nabla T_{e0}=-\widetilde{A_{j \parallel}} + \frac{C_T \alpha\alpha^\prime}{C_\nu e\nu_{ei}B_0}\nabla_{\parallel 0}\widetilde{T_e}.
    \label{eq: invert lapalce psi}
    \end{aligned}
\end{equation}

Electrostatic potential $\widetilde{\phi}$ and normalized parallel magnetic vector potential $\widetilde{\psi}$ are solved from Eqs. (\ref{eq: invert lapalce phi})-(\ref{eq: invert lapalce psi}) by the '\textit{invert-laplace}' Laplacian equation solver of BOUT++. This solver is introduced in detail in Appendix A. In our simulations, we exclude the current driven term by default, $B_0\times \nabla\widetilde{\psi}\cdot\nabla J_{\parallel 0}$, to focus on the temperature gradient driven MTM. 

\subsection{\label{sec:mode structure} Mode structure}

\begin{figure*}

    \centering
    \includegraphics[width=1.0\textwidth,keepaspectratio]{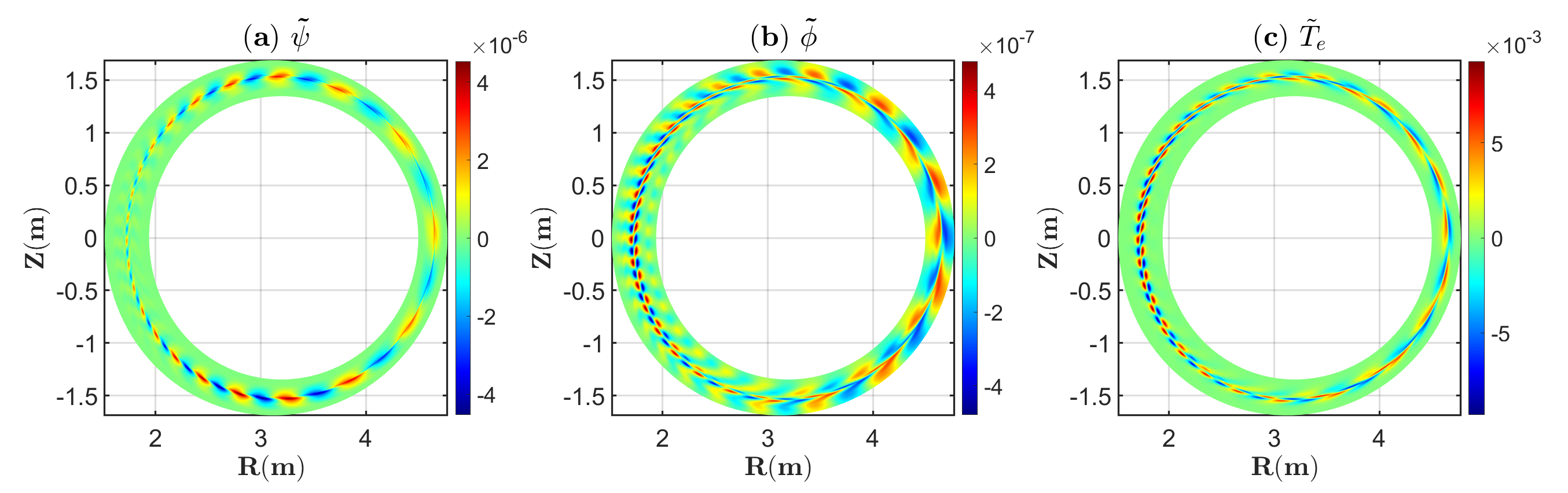}
    \includegraphics[width=0.95\textwidth,keepaspectratio]{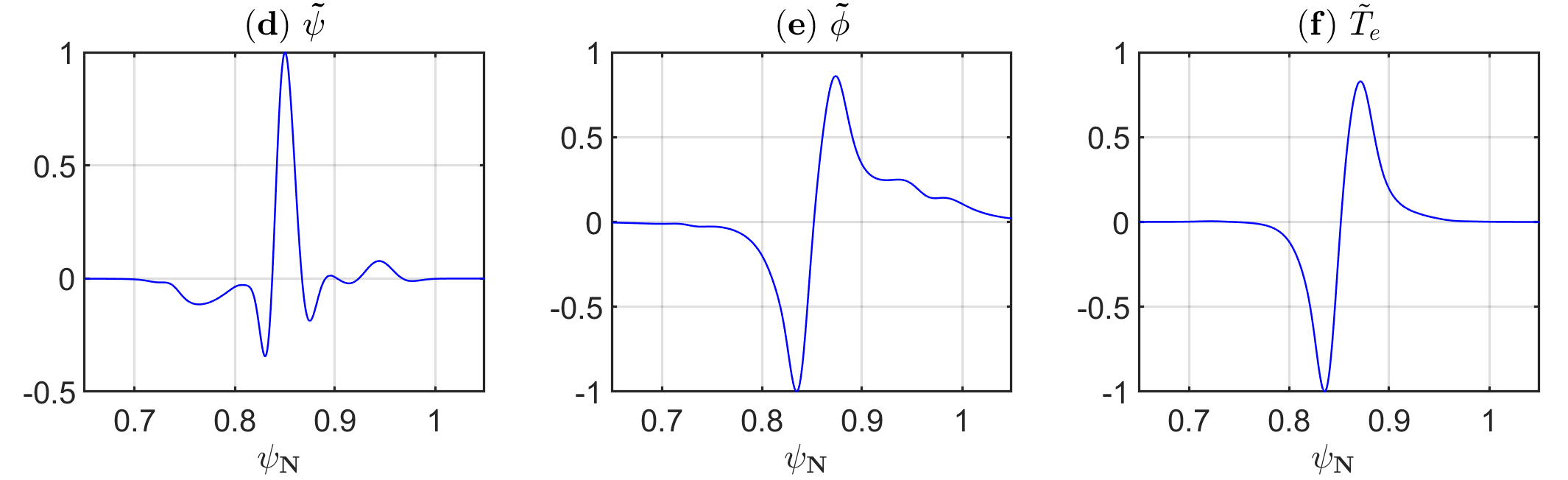}
   
    \caption{Poloidal distributions of (a) normalized parallel magnetic vector $\Tilde{\psi}$, (b) normalized electrostatic potential $\Tilde{\phi}$, and (c) normalized electron temperature perturbation $ \widetilde{T}_e$  for $n=8$ MTM. Radial distributions of (d) $\Tilde{\psi}$, (e) $\Tilde{\phi} $, and (f) $ \widetilde{T}_e$ at the inner mid-plane. $\widetilde{\psi}$ is normalized to $\Bar{L}$, $\widetilde{\phi}$ is normalized to $\Bar{B}\Bar{L}V_A$, and $\Tilde{T}_e$ is normalized to $\Bar{T}_e$.} 
    \label{fig:polslics phi and psi}

\end{figure*}

\begin{figure*}
    \centering
    \includegraphics[width=1.0\textwidth,keepaspectratio]{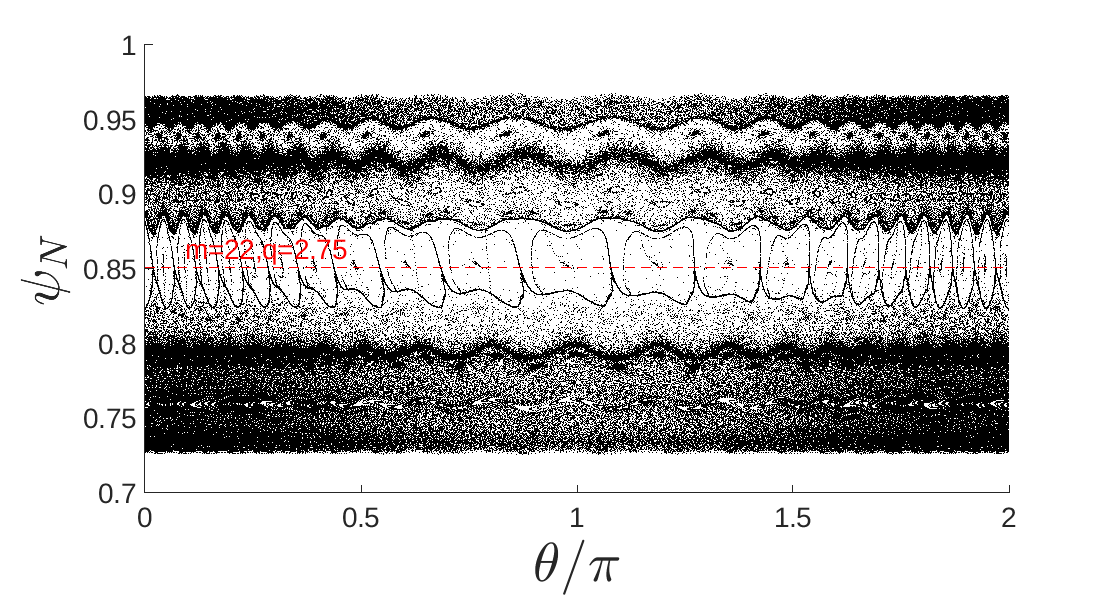}
    \caption{Poicare plot in a poloidal slice presents Magnetic island chains in the linear simulation at $t=800\tau_A$. The horizontal axis denotes the poloidal angle, where $\theta=\pi$ is the out mid-plane. The vertical axis denotes the normalized flux surface. The dashed red line indicates the location of $q=m/n=22/8=2.75$ rational surface.}
    \label{fig:magnetic island chains}
\end{figure*}

A linear simulation of $n=8$ is performed. FIG. \ref{fig:polslics phi and psi} displays the poloidal and radial distributions of $\Tilde{\psi}$, $\Tilde{\phi}$ and $\Tilde{T}_e$ at $t=800\tau_A$ in the linear simulation. These mode structures exhibit typical tearing parities, with $\Tilde{\psi}$ being even parity and $\Tilde{\phi}$ odd parity to the rational surface $q=2.75$. In FIG. \ref{fig:magnetic island chains}, a Poincaré plot reveals the formation of magnetic island chains around rational surfaces, providing compelling evidence for the identification of the MTM. 

\subsection{Dependency on collisionality and temperature gradient}\label{sec: gamma vs nu and grad Te} 
The dependence of linear growth rate and frequency on collisionality and temperature gradient is compared with theoretical expectations. In Sec.\ref{sec:dispersion MTM}, the conductivity is calculated, Eq. (\ref{eq: conductivity Hassam expression}). The lowest-order dispersion relation is employed, setting $\sigma(\omega)=0$, to derive the analytical expressions for growth rate and frequency,  

\begin{equation}
    \begin{aligned}
            \gamma\  &= \frac{(1+\alpha)C_T\omega_{*Te}}{1+(\alpha\alpha^\prime C_T\omega_{*Te}/C_\nu\nu_{ei})^2}\frac{\alpha\alpha^\prime C_T\omega_{*Te}}{C_\nu\nu_{ei}},\\
            \omega_r\  &= \frac{(1+\alpha)C_T\omega_{*Te}}{1+(\alpha\alpha^\prime C_T\omega_{*Te}/C_\nu\nu_{ei})^2}.
    \end{aligned}    
    \label{eq: gr and w by Hassam}
\end{equation}
This solution from the lowest order dispersion relation can capture the dependence trend. 

\begin{figure*}
    \centering
    \includegraphics[width=1.0\textwidth,keepaspectratio]{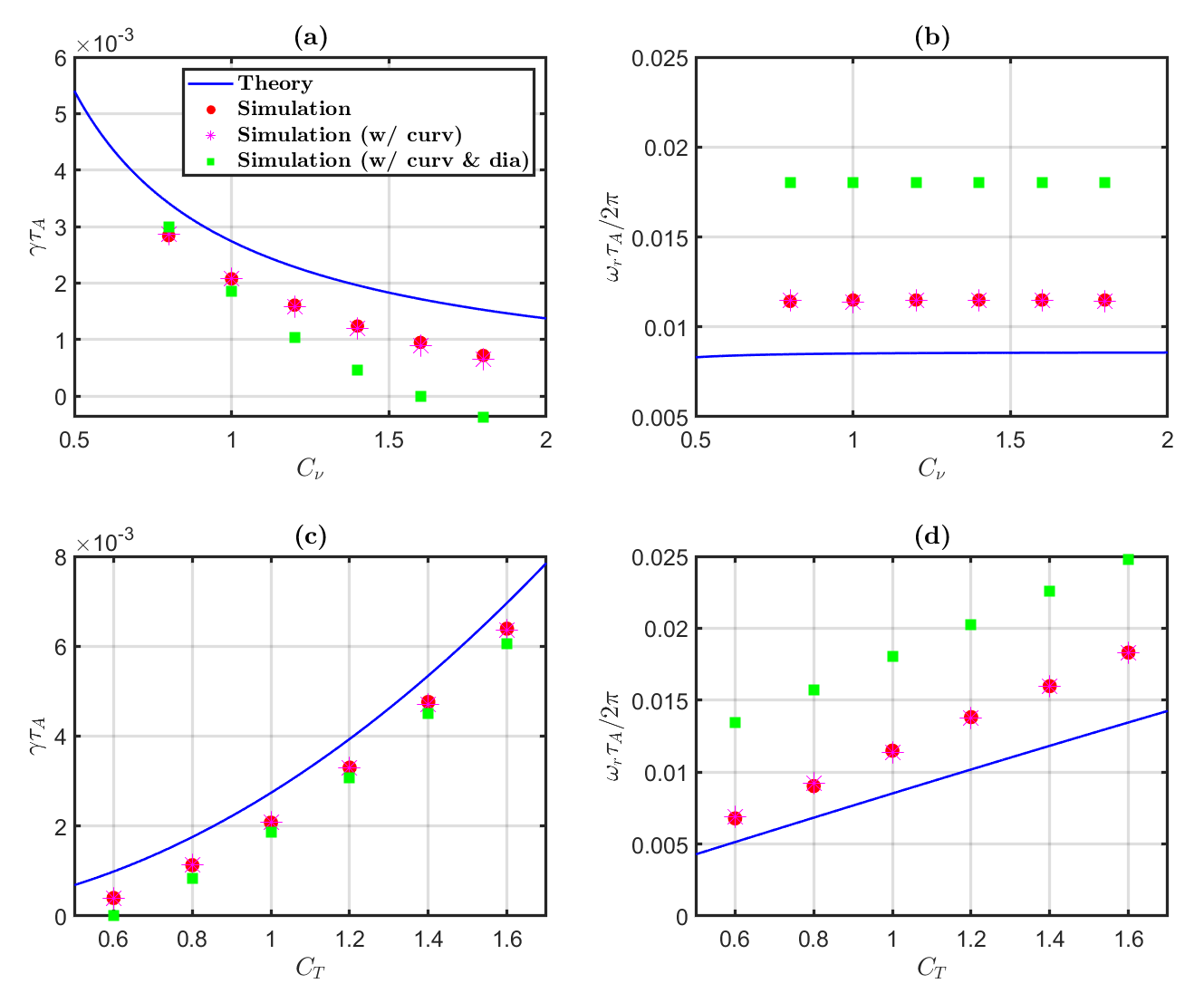}
    \caption{(a) Linear growth rate $\gamma$ and (b) mode frequency versus $C_{\nu}$ for $C_T=1.0$. The blue solid is calculated by Eq.(\ref{eq: gr and w by Hassam}); The scatter plots are the simulation results, where the red solid dots denote the simulations without curvature and diamagnetic drift effects, the magenta stars denote the simulations with curvature effect included, and the green squares denote the simulations with both curvature and diamagnetic drift effects included. (c) Linear growth rate and (d) mode frequency versus $C_{T}$ for $C_{\nu}=1.0$} 
    \label{fig:fr and gr vs Cnu}
\end{figure*}

We perform a scan in $C_{\nu}$ with $C_T=1.0$ for the $n=8$ mode to investigate the effects of collisionality. As shown in FIG. \ref{fig:fr and gr vs Cnu}(a), the growth rate decreases as collisionality increases. In comparison with the simulation, the local theory tends to overestimate the linear growth rate. This overestimation is attributed to the theory being local and not considering magnetic shear effects. The declining growth rate in the collisional regime with increasing collisionality is a distinct characteristic that sets it apart from the classical tearing mode. In FIG. \ref{fig:fr and gr vs Cnu}(b), it is evident that collisionality has a relatively minor impact on mode frequency in the collisional regime. The scan in $C_\nu$ indicates that collisionality plays a stabilizing effect for MTM. However, it is valid only in the strong collisional regime. MTM is destabilized by appropriate collisionality and temperature gradient. In the weak collision regime, the growth rate is proportional to collisionality, indicating that collisionality destabilizes MTM\cite{jian2019role} which distinguishes it from drift-wave instabilities. 

Another scan is performed in $C_T$ using the same approach, as presented in FIG. \ref{fig:fr and gr vs Cnu}(c). As $C_T$ increases, the growth rate shows a proportional increase. This is because, with the higher $C_T$, effectively both the thermal force and the time-dependent thermal force increase. These two forces drive the unstable MTM. In other words, when $C_T$ is higher, there is more effective free energy from the electron temperature gradient for electron motion and leads to greater instability in the MTM. This provides evidence that the MTM is primarily driven by the temperature gradient, as opposed to the current gradient characteristic for the classical tearing mode. The mode frequency is noticeably influenced by $C_T$, as shown in FIG. \ref{fig:fr and gr vs Cnu}(d). An increase in $C_T$ leads to a corresponding increase in the MTM frequency. This observation suggests that the MTM frequency is directly proportional to the electron diamagnetic frequency.

The effects of curvature and ion diamagnetic drift (correspondingly with the $E_r$ shear flow induced) are investigated. As illustrated in FIG.\ref{fig:fr and gr vs Cnu}, the magenta stars denote the cases with curvature drive (denoted by the third term on the RHS of Eq.(\ref{eq: linearized vorticity eq})) included and the green squares denote the cases with both curvature and ion diamagnetic drift effect included. The curvature almost has no influence on the linear growth rate and mode frequency indicating that the MTM is not a curvature-driven mode. 

However, when the ion diamagnetic drift effect is considered, the mode frequency and growth rate are well influenced. $E_r$ shear flow causes a Doppler shift in the frequency as discussed in the previous section and the change in mode frequency can be quantitatively illustrated. If performed in a slab geometry as in the Sec.\ref{sec:dispersion MTM} with $C_T=C_\nu=1.0$, the linearized electron momentum equation after the Fourier transform is, 
\begin{equation}
    \begin{aligned}
        \eta_s \widetilde{j_{\parallel}} \left(1- \frac{i\alpha^{\prime\prime}\omega}{0.51\nu_{ei}} \right) = \left[ 1-\left(1+\alpha+i\alpha\alpha^\prime\frac{\omega-\omega_{\phi_0}}{\nu_{ei}} \right) \frac{\omega_{*Te}}{\omega-\omega_{\phi_0}}  \right]\left[ i\left(\omega-\omega_{\phi_0}\right)\widetilde{A_{\parallel}}-ik_\parallel \widetilde{\phi}    \right]
    \end{aligned}
\end{equation}
where $\omega_{\phi_0}=k_y V_{E_0}=k_y\partial_x\Phi_0 / B_0$. Then we obtain a conductivity,
\begin{equation}
    \begin{aligned}
        \sigma_{dia} = \eta_s^{-1} \left[ 1-\left(1+\alpha+i\alpha\alpha^\prime\frac{\omega-\omega_{\phi_0}}{\nu_{ei}} \right) \frac{\omega_{*Te}}{\omega-\omega_{\phi_0}}  \right]   \left(1- \frac{i\alpha^{\prime\prime}\omega}{0.51\nu_{ei}} \right)^{-1}
    \end{aligned}
\end{equation}
Following a similar way, we can obtain the mode frequency and growth rate in the lowest order dispersion relation $\sigma_{dia}$,
\begin{equation}
    \begin{aligned}
            \gamma\  & \approx \frac{(1+\alpha)\omega_{*Te}}{1+(\alpha\alpha^\prime\omega_{*Te}/\nu_{ei})^2}\frac{\alpha\alpha^\prime\omega_{*Te}}{\nu_{ei}},\\
            \omega_r\  & \approx \frac{(1+\alpha)\omega_{*Te}}{1+(\alpha\alpha^\prime\omega_{*Te}/\nu_{ei})^2}+\omega_{\phi_0}.
    \end{aligned}    
    \label{eq: gr and w by Hassam dia}
\end{equation}
As illustrated in the previous section, the equilibrium $E_r$ flow balances the ion diamagnetic drift flow, $\mathbf{V}_{E_0}=\frac{\nabla P_i \times \mathbf{b}}{B_0}$. Correspondingly, $\omega_{\phi_0}=-\omega_{*i}=\omega_{*e}$. In the simulation, the density is constant, therefore, $\omega_{\phi_0}=\omega_{*Te}$. Approximately, we obtain the relation,

\begin{equation}
    \begin{aligned} 
     \frac{\omega_r|_{\textbf{w-dia}}}{\omega_r|_{\textbf{wo-dia}}} \approx \frac{2+\alpha+(\alpha\alpha^\prime\omega_{*Te}/\nu_{ei})^2}{1+\alpha}\approx 1.56.
    \label{eq: gr and w : dia vs modia}
    \end{aligned}
\end{equation}
This relation indicates that considering the diamagnetic drift effect, along with the inclusion of equilibrium $E_r$ shear flow, results in a mode frequency approximately 1.56 times higher than when the diamagnetic effect is not considered. This relation aligns with the simulation results displayed in FIG. \ref{fig:fr and gr vs Cnu}(b) and (d) when $C_T,\ C_\nu=1.0$.

\begin{figure*}
    \centering

    \includegraphics[width=0.6\textwidth,keepaspectratio]{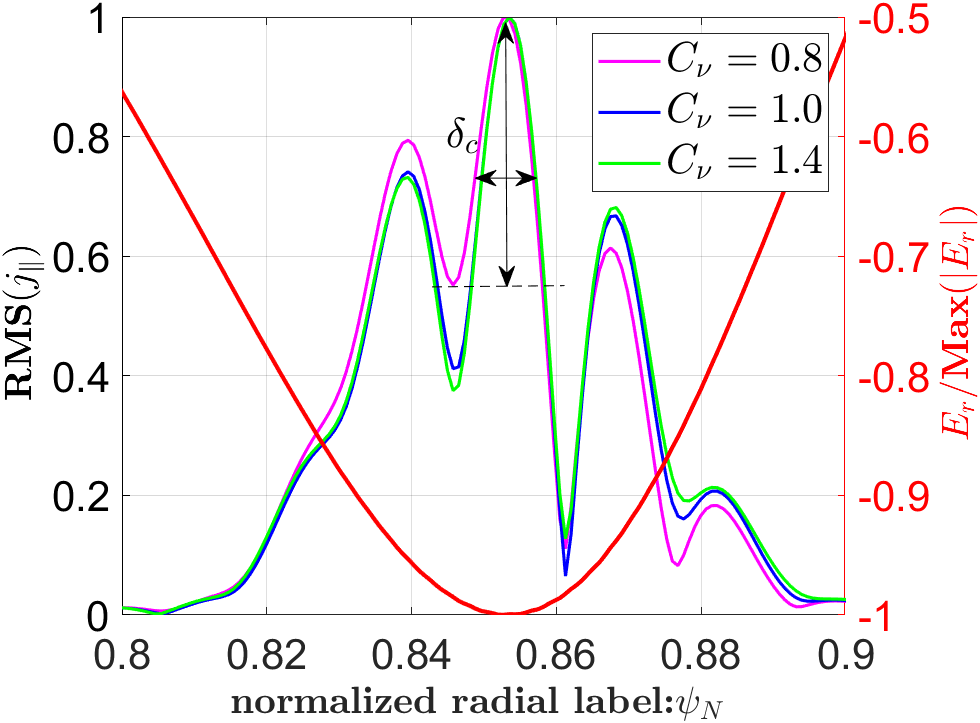}
    \caption{Diagram of $E_r$ profile and distributions of RMS $j_\parallel$ at outer mid-plane of $n=8$ modes for $C_\nu=0.8$, $C_\nu=1.0$, and $C_\nu=1.4$ with $C_T=1.0$. } 
    \label{fig: Er and mdoe structure Cnu1.0 and Cnu1.4}
\end{figure*}

The reduction in growth rate is attributed to the non-local shear of $E\times B$ flow. As illustrated in FIG.\ref{fig: Er and mdoe structure Cnu1.0 and Cnu1.4}, RMS $j_\parallel$ distributions at outer mid-plan of different collisonality are presented. Following the definition of Yagyu and Numata\cite{yagyu2023destabilization}, the current layer width $\delta_c$ is defined by the full width at half maximum height of $j_\parallel$ where the height is measured from the top to the first local minimal value. The current layer widths of $C_\nu=0.8,1.0,1.4$ are $\delta_c /\rho_i=3.22,3.5,3.63$. MTM is localized around the bottom of the $E_r$ well where the local shear is nearly zeros. In such instances, the $E_r$ shear flow has minimal impact on the linear growth rate. However, as collisionality increases, the current layer width broadens. Consequently, the $E_r$ shear flow could suppress the MTM leading to a decrease in the linear growth rate.     

\subsection{Alignment of rational surfaces with $\omega_{*e}$ profile}\label{sec:alignment effect}
MTM is unstable only when the rational surface is in good alignment with the peak of the electron diamagnetic frequency, $\omega_{*e}$, profile. As pointed out by Larakers and Curie\cite{curie2022survey}, MTM is unstable with $\Delta_s/x_*<0.3$, where $\Delta_s$ is the distance that one rational surface shifts away from the peak of the location of $\omega_{*e}$ and $x_*$ is the characteristic width of $\omega_{*e}$ profile. Once the rational surface is too far away from the peak $\omega_{*e}$, MTM can not survive. The competition between the nonalignment damping effect and gradient-driven effect determines the unstable regime of MTM. And the nonalignment is a stabilizing effect. Our conclusion is consistent with Larakers' global calculation\cite{larakers2021global}, Curie's SLiM model\cite{curie2022survey,curie2023microtearding}, and also Hassan's pedestal simulation\cite{hassan2021identifying}.

\begin{figure*}
    \centering
    \includegraphics[width=0.9\textwidth,keepaspectratio]{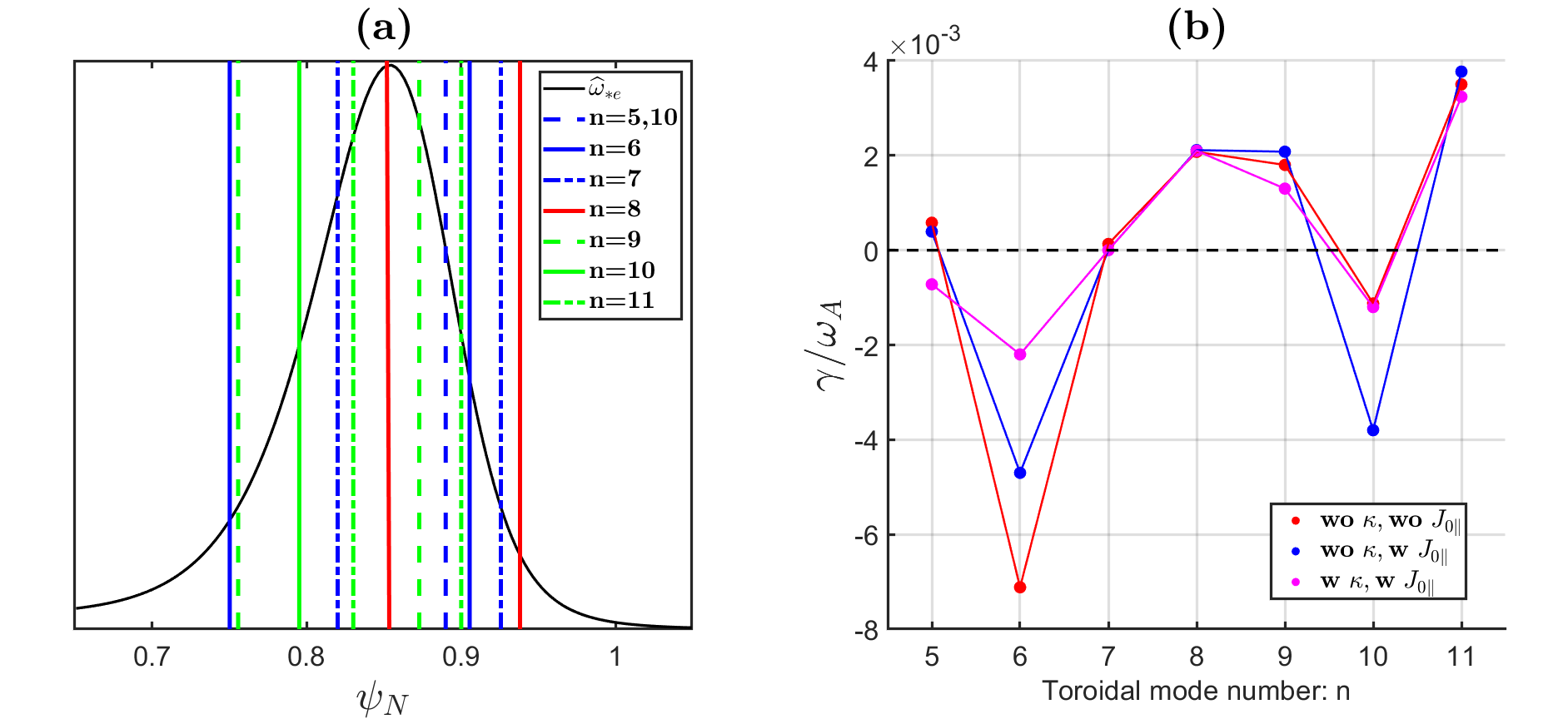}
     \includegraphics[width=1.0\textwidth,keepaspectratio]{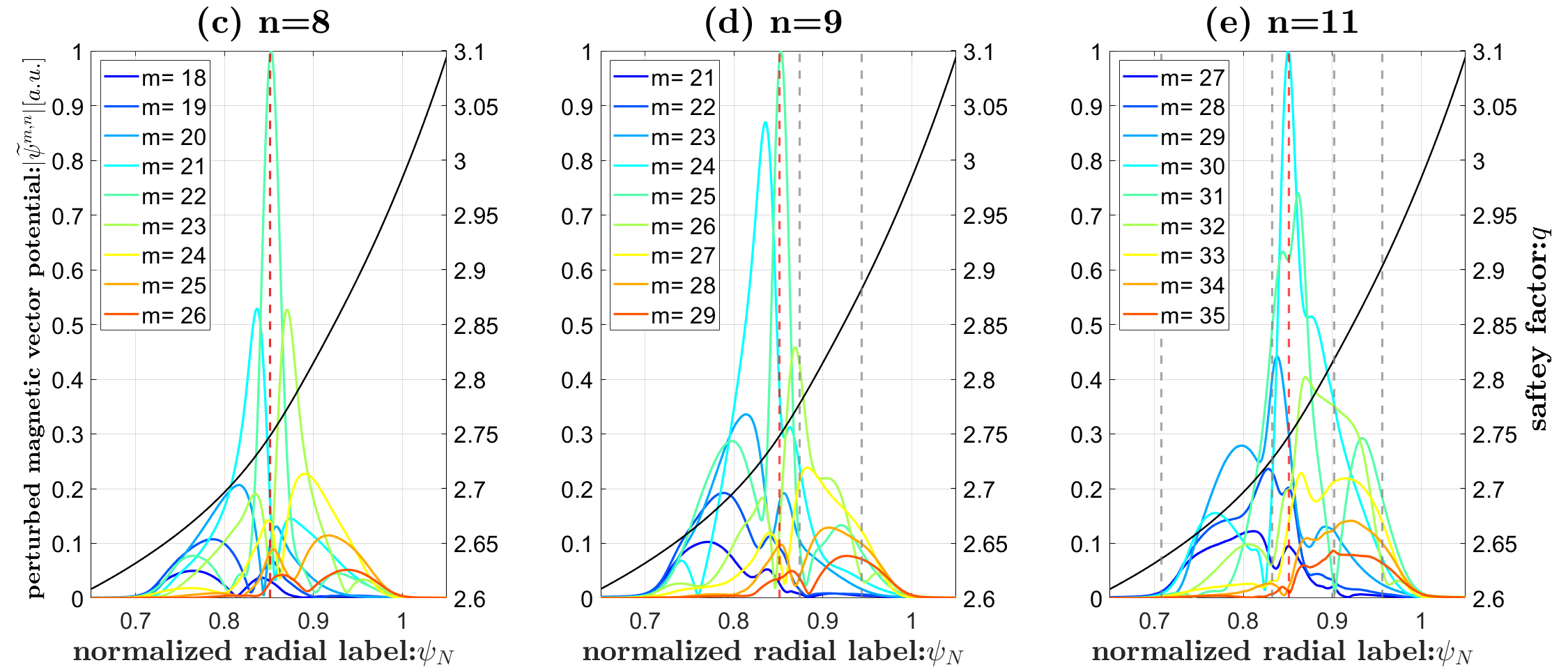}
    
    \caption{(a) The diagram of $\omega_{*e}$ profile and location of rational surfaces of different toroidal mode numbers; (b) the growth rate of different toroidal mode number cases. (c)-(e) presents the mode structure of different poloidal number modes corresponding to $n=8,9,11$ cases. The dashed vertical red line denotes the location of $\omega_{*e}$ peak and the the dashed vertical grey lines denote the location of rational surfaces. The solid black curve represents the safety factor profile.}
    \label{fig:rational surfaces and gr spectrum}
\end{figure*}

A linear simulation is conducted to confirm this prediction. The equilibrium temperature and safety factor profiles align with those shown in FIG. (\ref{fig:Eqprofile}). Toroidal modes are scanned from $n=5$ to $n=11$. In FIG. \ref{fig:rational surfaces and gr spectrum}(a), the locations of rational surfaces for different toroidal modes are labeled with various colored vertical lines. Correspondingly, FIG. \ref{fig:rational surfaces and gr spectrum}(b) displays the growth rate spectrum. The red dotted line denotes the cases without curvature effect and peeling drive effect (related to $J_{\parallel 0}$ and denoted by the second term on the RHS of Eq.(\ref{eq: linearized vorticity eq})), the blue dotted line denotes the cases with peeling drive effect included, and the magenta line denotes the cases with both curvature and peeling drive effect included.

The rational surface of $n=8$ almost perfectly aligns with the peak of $\omega_{*e}$, while the rational surface of $n=9$ is slightly shifted. Under the discussion in the previous section, when the drive increases, the growth rate is correspondingly higher. And a higher toroidal mode number leads to a more pronounced $\omega_{*e}$ profile since $\omega_{*e} \propto n$, and then increases the drive. The mode structures of the most unstable mode, $n=8,9,11$, in Fig.\ref{fig:rational surfaces and gr spectrum}(b) are presented in Fig.\ref{fig:rational surfaces and gr spectrum}(c)-(e). Our global simulations show that the perturbations peak at the location of peak $\omega_{*e}$ but not always at the rational surface. For $n=9,11$ mode, the most unstable $m$ modes offset the rational surface and the symmetry of the mode structure breaks down. As predicted by the local theory, the $n=9$ mode is expected to have a larger growth rate than that of $n=8$. However, $n=9$ has a smaller growth rate despite having a larger $\omega_{*e}$. This is due to the non-local nonalignment effect, the linear growth rate of the $n=9$ mode decreases. 

As the blue dotted line in FIG.\ref{fig:rational surfaces and gr spectrum}(b) shows, $J_{0\parallel}$ destabilizes the non-aligned modes, like $n=9$ mode, but has little influence on the aligned $n=8$ mode. The curvature effect can stabilize the non-aligned mode as the magenta dotted line shows in FIG.\ref{fig:rational surfaces and gr spectrum}(b).

 

\section{Global nonlinear simulation of MTM}\label{sec:nonlinear global simu}
In this section, we conduct an initial nonlinear simulation to explore the nonlinear saturation mechanism of MTM. We will compare radial thermal transport resulting from $ {E}\times  {B}$ convection and magnetic flutter. The nonlinear simulation retains all terms except the peeling drive in Eqs.(\ref{eq: vorticity evolution})-(\ref{eq: temperature evolution}) and utilizes the same equilibrium as in the previous section. Also, the $E_r$ shear flow is included by default in the following simulations.

Due to the limitations of the collisional Hassam model, accurate calculations of high-$n$ modes are not feasible. Therefore, the nonlinear simulation includes only $n=4,8,12$ modes while retaining the zonal component of temperature perturbation. As the rational surfaces for these three modes align with the peak $\omega_{*e}$, the nonalignment stabilizing effect is excluded. The initial perturbation is magnetic and takes the form of,
\begin{equation}
    \widetilde{\psi}_{ini}=\exp{\left[-\left(\frac{x-x_0}{\delta_x}\right)^2-\left(\frac{y-y_0}{\delta_y}\right)^2\right]}\sum_{n=1}^{3} A_n \cos{(nk_z+\varphi_n)}
\end{equation}
where $\delta_x=0.1,\delta_y=0.3,x_0=0.5,y_0=0.5$, $k_z=2\pi/4$, $\varphi_n$ is a random phase. 

\begin{figure*}
    \centering
    \includegraphics[width=0.9\textwidth,keepaspectratio]{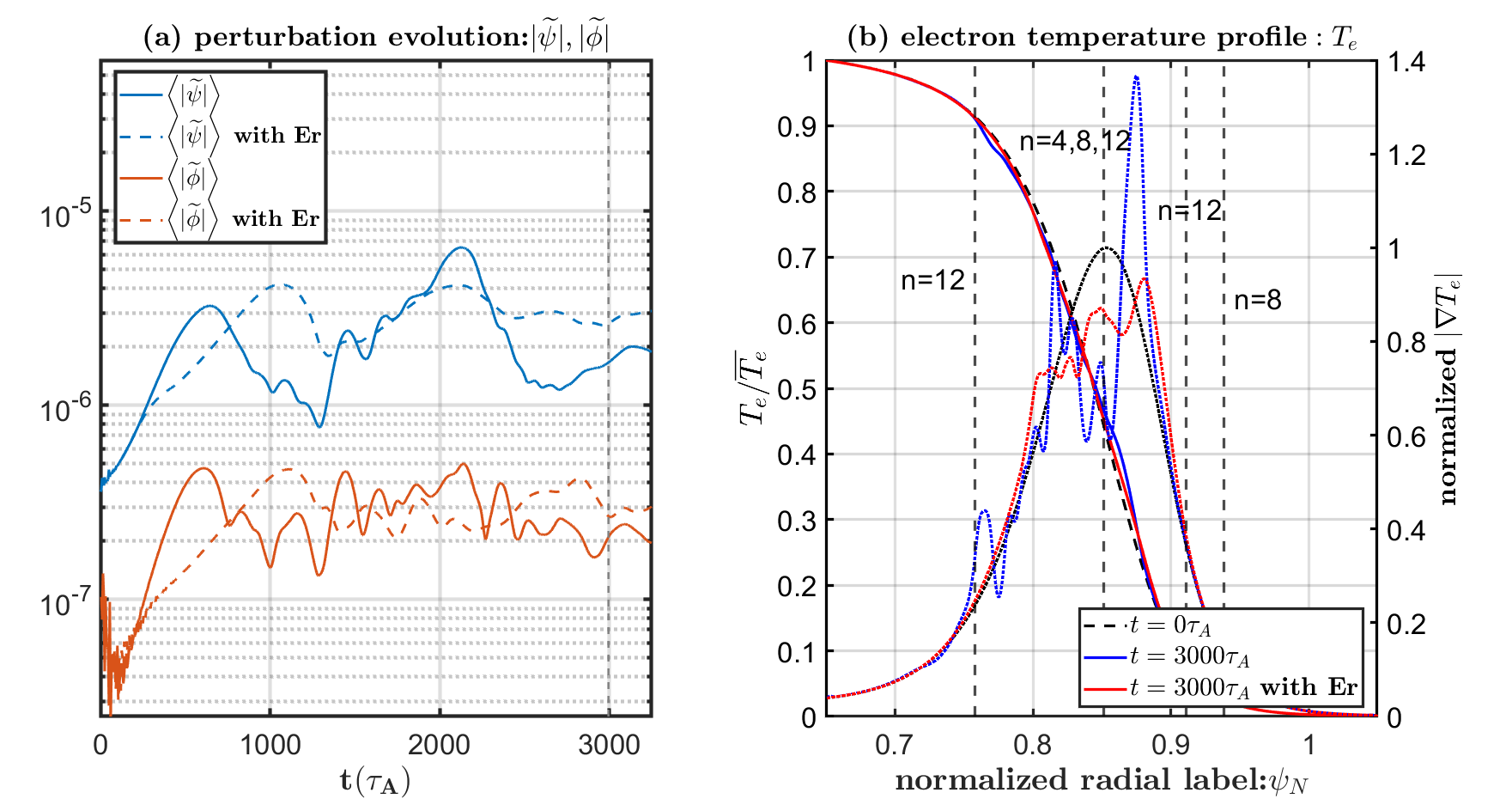}
    \caption{(a) The time evolution of the root-mean-square (RMS) of magnetic perturbations is depicted. The solid blue and orange curves represent the evolution of flux surface-averaged magnetic vector potential and electrostatic potential perturbations at the $q=2.75$ rational surface, respectively. The dashed curves indicate the perturbation evolution with equilibrium $E_r$ shear flow. (b) The electron temperature profiles at different time are illustrated. The dashed black curve represents the initial profile, the blue curve represents the profile after $3000\tau_A$, and the red curve represents the profile after $3000\tau_A$ with equilibrium $E_r$ shear flow. The corresponding normalized temperature gradient profiles are plotted in dotted lines with the same color.}
    \label{fig:RMS Psi and profile change}
\end{figure*}

\subsection{Saturation of collisional MTM due to profile relaxation}\label{sec:saturation mach}

The surface-averaged perturbations of the absolute values of electrostatic potential $\widetilde{\phi}$ and normalized parallel magnetic vector potential $\widetilde{\psi}$ are presented in FIG. \ref{fig:RMS Psi and profile change}(a). After approximately $1000\tau_A$ time, the MTM enters a saturation state. Notably, the temperature profile changes the evolution. We depict the initial temperature profile with a dashed black curve, and the profile after 3000 Alfvén times with a solid blue curve, as shown in FIG. \ref{fig:RMS Psi and profile change}(b). The corresponding normalized gradient profiles are shown as dotted lines with the same colors.

It is evident that the temperature profile changes, with the peak gradient decreasing occurring around the $q=2.75$ rational surface. Consequently, the linear-driven force decreases, leading to a gradual cessation of perturbation growth, ultimately resulting in mode saturation. In our nonlinear simulations, Only the zonal component of temperature perturbation $\Tilde{T}_e$ is kept, the zonal component of $\widetilde{\phi}$ and $\widetilde{\psi}$ are removed, and correspondingly, zonal flow and zonal field are not included. We are not going to discuss how they affect the saturation process and will discuss it in another paper. In comparison with the scenario without the equilibrium $E_r$ shear flow, perturbations exhibit a reduced linear growth rate but maintain a similar amplitude in the nonlinear phase. Without the equilibrium $E_r$ shear flow, the temperature gradient exhibits a more significant decrease.

\begin{figure*}
    \centering
    \includegraphics[width=1.0\textwidth,keepaspectratio]{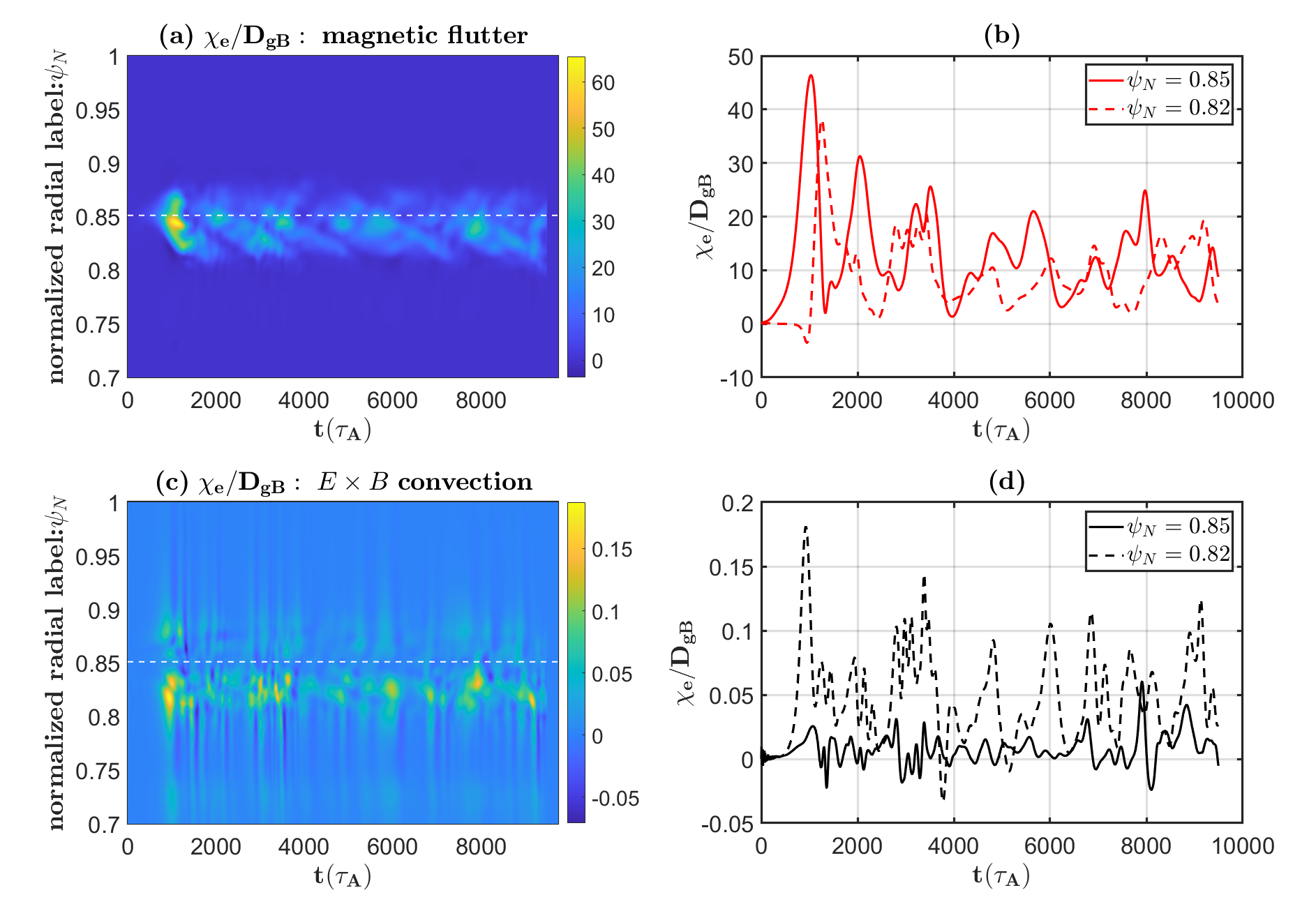}
  
    \caption{The surface averaged thermal transport coefficient is normalized by the Bohm diffusion coefficient. (a) and (c) represent the temporal evolution of radial distributions of magnetic flutter component and $E\times B$ convection respectively. The horizontal dashed white line denotes the evolution of $ \chi_e $ at $q=2.75, \ \psi_N=0.85$ rational surface. Correspondingly, they are plotted in solid lines in panels (b) and (d).}
    \label{fig:heat flux}
\end{figure*} 

\begin{figure*}
    \centering
    \includegraphics[width=0.9\textwidth,keepaspectratio]{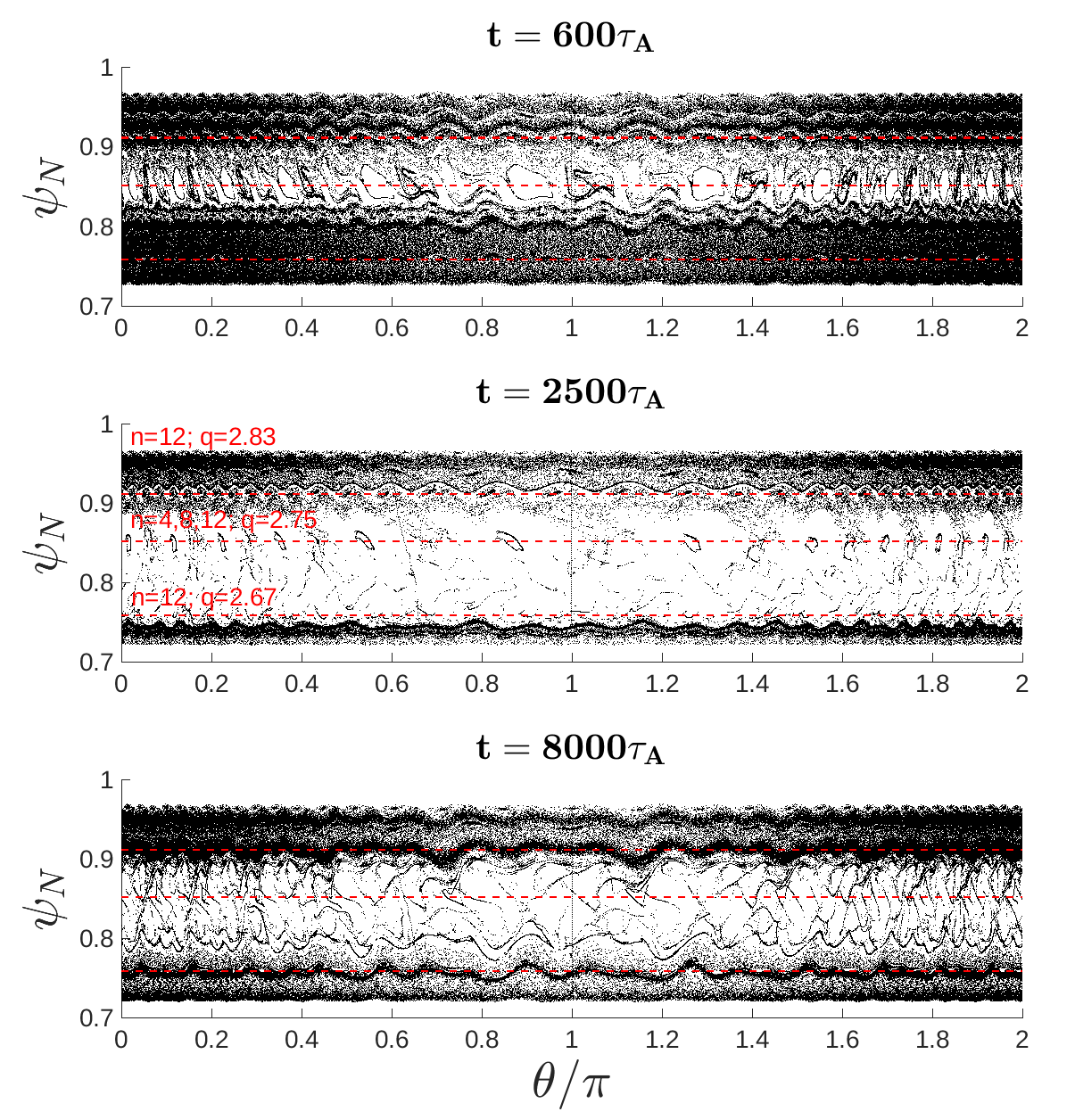}
    \caption{Poincare plots in the nonlinear simulation at $t=600\tau_A$ (top), $t=2500\tau_A$ (middle), and $t=8000\tau_A$ (bottom), corresponding to the simulation case with $E_r$ shear flow. The horizontal dashed red lines show the location of different rational surfaces.}
    \label{fig: Poincare plot nonlienar simu}
\end{figure*}

\subsection{Electron radial thermal transport}\label{sec: thermal transport}
MTM is one candidate for anomalous electron thermal transport in the H-mode discharge. The electron radial thermal transport can be divided into two parts, one is the $E \times B$ convection part and the other is the magnetic flutter part. The total heat flux has a form\cite{xia2015nonlinear},
\begin{equation}
\begin{aligned}
    \left<Q_{er}\right> =  \left<V_E p_e \right> + \left<  -\frac{B_r}{B_0} \kappa_{fl} \nabla_{\parallel} T_e \right>
    \label{eq: heat flux}
\end{aligned}
\end{equation}
where $p_e\equiv n_e T_e$ is the electron pressure, $B_r = \bm{b_0}\times \nabla \widetilde{\psi}$ is the radial component of the perturbed magnetic field, the bracket $\left<...\right>$ denotes the magnetic surface average via the unperturbed surfaces. The first term at the right-hand side of Eq.(\ref{eq: heat flux}) is the $E\times B$ convection heat flux and the second term is the magnetic flutter heat flux. A surface-averaged thermal transport coefficient is defined as,
\begin{equation}
    \begin{aligned}
        \chi_e = \left< \frac{Q_{er}}{-n_{e}\nabla T_{e}}  \right>. 
    \label{eq: chie defination}
    \end{aligned}
\end{equation}

To illustrate the MTM thermal transport, we use the case with $E_r$ shear flow as an example. The normalized $ \chi_e $ of the two components are shown in FIG.\ref{fig:heat flux}. The normalized factor $D_{gB}= {T_e}\rho_i^*/16e {B_0}=0.0071[\text{m}^2\cdot \text{s}^{-1}])$ represents the gyro-Bohm diffusion coefficient, and $\rho_i^*$ is the ion gyroradius normalized to minor radius. FIG.\ref{fig:heat flux}(a) and (c) present the temporal evolution of the distributions of the two components of $ \chi_e $. In comparison with the magnetic flutter component, the electron thermal transport by $E\times B$ convection is negligible. This is attributed to the changes in magnetic field topology induced by MTM. It can be illustrated in FIG.\ref{fig: Poincare plot nonlienar simu}. In the nonlinear phase, such as $t=2500\tau_A$, the original closed flux surfaces break down due to the strong magnetic perturbation compared to that in the linear phase at $t=600\tau_A$. There forms a magnetic island chain around the $q=2.75$ rational surface and a stochastic magnetic field in the major domain between $q=2.67$ and $q=2.833$ rational surfaces. For the electrons, the parallel conductivity is large, and the parallel thermal transport has a radial component due to the magnetic perturbation. Therefore, the electron thermal transport is enhanced, and the magnetic flutter component dominates the electron thermal transport.      

In our simulation, we observe an asymmetric spreading of MTM turbulence. There is an evident inward spreading as indicated in FIG.\ref{fig:heat flux}(a). Such an inward spreading intensifies the thermal transport near the pedestal top. As shown in FIG.\ref{fig:heat flux}(b),  the thermal transport coefficient at $\psi_N=0.82$ increases after a temporal delay and eventually reaches the same level as that at $\psi_N=0.85$.

\subsection{The influence of the free streaming parameter $\alpha_H$ on electron radial thermal transport}\label{sec:paramter aq}
In this section, we discuss how the free streaming parameter $\alpha_H$ in the flux limit heat flux model affects the electron thermal transport in the global nonlinear simulations. 

\begin{figure*}
    \centering
    \includegraphics[width=1.0\textwidth,keepaspectratio]{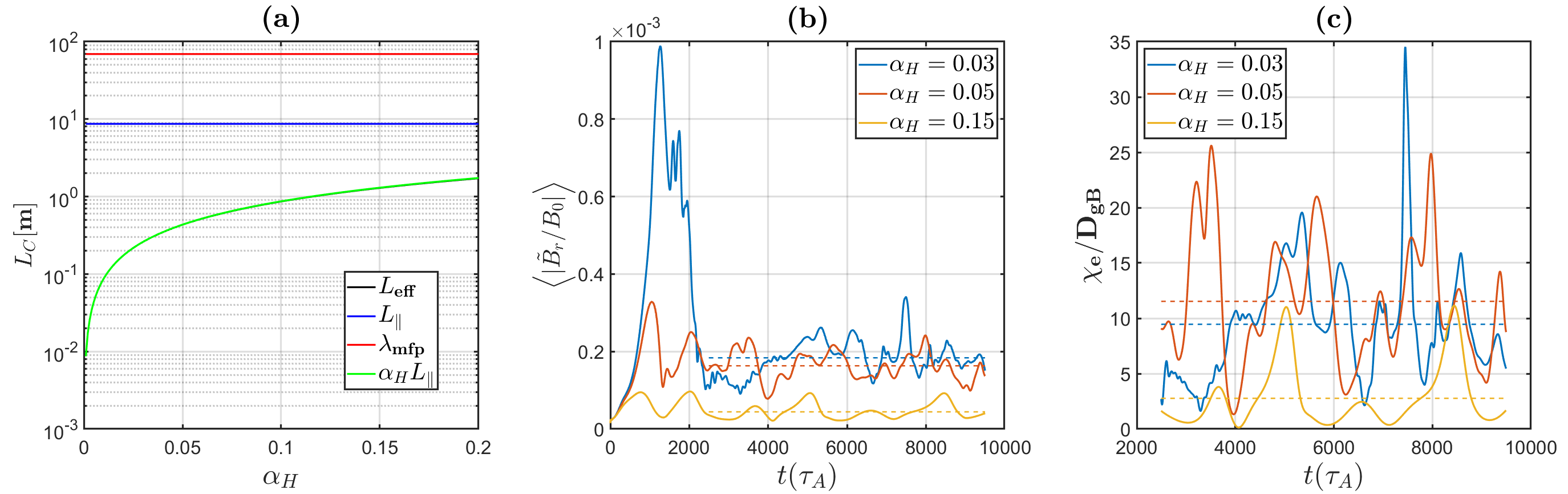}

    \caption{(a)Total effective correlation length $L_{\text{eff}}$, parallel correlation length $L_\parallel $, collision mean free path $\lambda_{\text{mfp}}$, and parallel effective correlation length $\alpha_H L_\parallel$ versus $\alpha_H$; (b) Temporal evolution of flux surface averaged $\Tilde{B}_r$ at $q=2.75$ rational surface. The dashed horizontal lines denote the period-averaged $\left<|\Tilde{B}_r/B_0|\right>$ within $t=3000-9500\tau_A$. The averaged $\left<|\Tilde{B}_r/B_0|\right>$ are $1.84\times 10^{-4}$, $1.63\times 10^{-4}$, $4.43\times 10^{-5}$ for $\alpha_H=0.03,\ 0.05,\ 0.15$. (c) The electron thermal transport coefficient of the three cases within the nonlinear phase $t=3000-9500\tau_A$. The averaged values are denoted by the dashed horizontal lines and they are 9.45, 11.53, and 2.77 normalized to $D_{gB}$ for $\alpha_H=0.03,\ 0.05,\ 0.15$ respectively. }
    \label{fig: lengths comparison and perturbation}
\end{figure*}

The magnetic flutter component dominates electron thermal transport. This component can be written in an explicit form with perturbation and equilibrium parts separated. 
\begin{equation}
    \begin{aligned}
        Q_{er}^{\text{mf}} =  \frac{ \Tilde{q}_{e\parallel} \Tilde{B}_r}{B_0} = -n_e\kappa_{fl} \left[ \frac{\Tilde{B}_r}{B_0}\nabla_{\parallel 0} \Tilde{T}_e  + \frac{\Tilde{B}_r^2}{B_0^2}\nabla \left( T_{e0} + \Tilde{T}_e |_{\text{DC}} \right) + \frac{\Tilde{B}_r^2}{B_0^2}\nabla \Tilde{T}_e|_{\text{AC}} \right].
        \label{eq: magnetic flutter Qer}
    \end{aligned}
\end{equation}
Here, the subscript \textit{DC} denotes the zonal component, and \textit{AC} denotes the non-zonal component. $\Tilde{T}_e |_{\text{DC}}$ change the electron temperature profile, $\nabla \left( T_{e0} + \Tilde{T}_e |_{\text{DC}} \right)\equiv \nabla T_e $ is the electron temperature gradient at the moment. The first term in the bracket of Eq.(\ref{eq: magnetic flutter Qer})  describes the parallel dissipation and decreases the magnetic flutter electron thermal transport. The second term dominates the radial thermal transport, it describes an effective radial thermal diffusion when the magnetic filed is perturbed. The third term has little contribution to the heat flux after the magnetic surface average. A magnetic flutter thermal transport coefficient can be defined,
\begin{equation}
    \begin{aligned}
        \chi_e^{\text{mf}} \equiv \left< \frac{Q_{er}^{\text{mf}}}{-n_e\nabla T_e}\right>. 
    \end{aligned}
\end{equation}
Approximately, $\chi_e = \chi_e^{\text{mf}}$. The flux-limit $\kappa_{fl}$ in Eq.(\ref{heatflux}) can be rewritten in a form,
\begin{equation}
    \begin{aligned}
        \kappa_{fl} =n_e v_{te} / \left[ (\alpha_H L_\parallel)^{-1} + (3.2 \lambda_{\text{mfp}})^{-1}\right],
    \end{aligned}
\end{equation}
where $\alpha_H L_\parallel$ is an effective correlation length along the magnetic field line, and $\lambda_{\text{mfp}}= v_{te}/\nu_{ei}$ is the collision mean free path describing a decorrelation length due to collision. Then, a total effective correlation length is defined,
\begin{equation}
    L_{\text{eff}} = \frac{1}{(\alpha_H L_\parallel)^{-1} + (3.2 \lambda_{\text{mfp}})^{-1}}.
\end{equation}
Using the equilibrium parameters, $L_{\text{eff}}$ is calculated and presented in FIG.\ref{fig: lengths comparison and perturbation}(a). $\alpha_H L_{\parallel}$ is comparable to the device size and much smaller than the collision mean free path and determines the $L_{\text{eff}}$. Therefore, $L_{\text{eff}}$ is roughly proportional to the free streaming parameter $\alpha_H$.

Because the second term in the bracket in Eq.(\ref{eq: magnetic flutter Qer}) dominates the $Q_{er}^{\text{mf}}\propto \chi_e$, electron thermal transport can be rewritten as 
\begin{equation}
    \chi_e^{\text{mf}} \approx C_{\text{mf}}\left<v_{te}L_{\text{eff}}(\alpha_H) \frac{\Tilde{B}_r^2}{B_0^2}\right>,
    \label{eq: chie vte Leff Br2}
\end{equation}
with coefficient $C_{\text{mf}}$ is less than unity.

The electron parallel heat flux $q_{e\parallel}$ plays two roles. It doesn't only contribute to a large thermal transport but also dissipates the perturbations. As $\alpha_H$ increases, $\chi_{fl} \equiv v_{te}L_{\text{eff}}(\alpha_H)$ increases correspondingly. But a larger $\chi_{fl}$ makes perturbation, like $\Tilde{B}_r$, better damped as shown in FIG.\ref{fig: lengths comparison and perturbation}(b). An increase in $\alpha_H$ from 0.03 to 0.15 makes the perturbation $\left<\Tilde{B}_r/B_0\right>$ decrease from $1.84\times 10^{-4}$ to $4.43\times 10^{-5}$. Therefore, these two effects compete to determine electron thermal transport. As shown in FIG.\ref{fig: lengths comparison and perturbation}(c), when $\alpha_H=0.05$, the averaged electron thermal transport coefficient is the largest in comparison with that of $\alpha_H=0.03$ and $\alpha_H=0.15$.

\begin{figure*}
    \centering
    \includegraphics[width=1.0\textwidth,keepaspectratio]{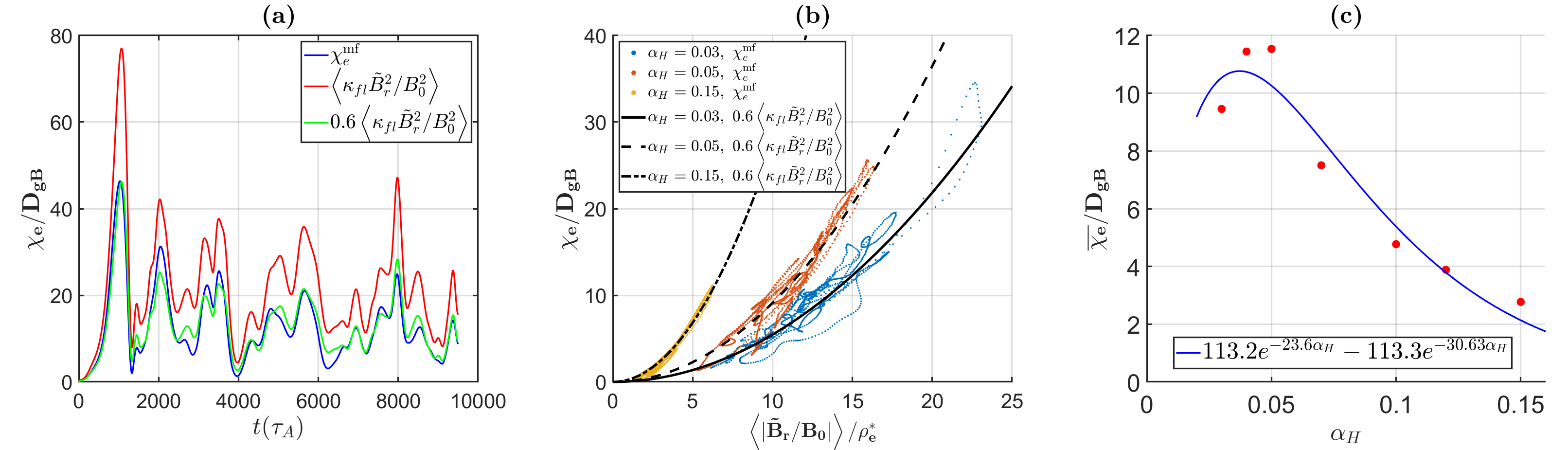}

    \caption{At the $q=2.75$ rational surface, (a) temporal evolution of total magnetic flutter electron thermal transport coefficient (blue),  $\chi_e^{\text{mf}}$, for $\alpha_H=0.05$; (b) The dots in different colors denote the $\chi_e^{\text{mf}}$ calculated at each time-step of different $\alpha_H$ cases within $t=2500-9500\tau_A$. Black curves in different styles are calculated from Eq.(\ref{eq: chie vte Leff Br2}). $\rho_e^*$ is the electron gyro-radius normalized to minor radius; (c) Time averaged thermal transport coefficient (red dots) versus the free streaming parameter $\alpha_H$.}
    \label{fig: chie vs time and chie vs Bx}
\end{figure*}

We confirm the value of coefficient $C_{\text{mf}}$ to be 0.6 in the simulations. As shown in FIG.\ref{fig: chie vs time and chie vs Bx}(a), if choosing $C_{\text{mf}}=0.6$, $C_{\text{mf}}\left<v_{te}L_{\text{eff}}(\alpha_H) {\Tilde{B}_r^2}/{B_0^2}\right>$ is well quantitatively fitted with $\chi_e^{\text{mf}}$. Not only for the case $\alpha_H=0.05$ but also for the other two cases as shown in FIG.\ref{fig: chie vs time and chie vs Bx}(b). We display the results on the $|\Tilde{B}_r| - \chi_e$ plane and every single dot denotes one result at one time-step. $C_{\text{mf}}=0.6 $ indicates that the parallel dissipation makes a 40\% decrease in the magnetic flutter component of electron thermal transport. In our simulation, we find out that for different $\alpha_H$, the electron thermal transport follows the same scaling law,
\begin{equation}
    \chi_e \approx 0.6\left< \kappa_{fl} \frac{\Tilde{B}_r^2}{B_0^2}\right>.
    \label{eq: chie 0.6 chifl Bx}
\end{equation}
A thermal transport coefficient in this form indicates that thermal transport can be conducted by a stochastic magnetic field \cite{rechester1978electron,doerk2012gyrokinetic}. It is consistent with the discussion in Sec.\ref{sec: thermal transport}.

Simulations for more different $\alpha_H$ are performed and the time averaged thermal transport coefficients $\overline{\chi_e}$ are presented in FIG.\ref{fig: chie vs time and chie vs Bx}(c). The maximum of $\overline{\chi_e}$ appears in the regime of $\alpha_H \in [0.04,\ 0.05]$. A fitted scaling law as indicated by the blue curve in FIG.\ref{fig: chie vs time and chie vs Bx}(c) is,
\begin{equation}
    \begin{aligned}
        \overline{\chi_e} = \left(113.2e^{-23.6\alpha_H}-113.3e^{-30.63\alpha_H}  \right) D_{gB}
    \end{aligned}
\end{equation}

However, from the discussion above, we learn that electron thermal transport is strongly dependent on the model of parallel heat flux. Our flux-limit heat flux model can capture some instinct characteristics of microtearing turbulence like the transport scaling law Eq.(\ref{eq: chie 0.6 chifl Bx}), but quantitatively the results are determined by the  $\alpha_H$ we chose.

\section{Summary}\label{sec:summary}
In our research, we have developed a fluid simulation code for studying the collisional MTM within the BOUT++ framework. We should note that BOUT++ has been updated to include the capability for simulating this mode and make it possible to investigate how MTM interacts with other ion modes like the kinetic ballooning mode. 

In our linear simulations, we observe that the MTM exhibits a tearing parity in its mode structure. We explore the dependencies of growth rate and mode frequency on collisionality and temperature gradient. Within the collisional regime, the linear growth rate of the MTM decreases as collisionality increases, while it remains directly proportional to the electron temperature gradient. The $E_r$ shear flow causes a reduction in the linear growth rate especially when the collision is strong. Additionally, we investigate the non-local alignment effect, revealing that the MTM becomes unstable only when the rational surface aligns with the peak of the $\omega_{*e}$ profile; and the nonalignment effect serves as a stabilizing factor. 

In our nonlinear simulations, we find that the saturation of the MTM is primarily due to electron temperature profile relaxation with decreasing temperature gradient around the rational surface. Furthermore, we observe that electron thermal transport from magnetic flutter significantly exceeds that generated by $\bm{E}\times \bm{B}$ convection due to the change in the topology of the magnetic field by MTM. Evident inward turbulence spreading is observed to intensify the thermal transport near the pedestal top. In our simulations, the electron thermal transport is quantitatively influenced by the heat flux closure or the free streaming parameter $\alpha_H$ in the flux-limit heat flux model. 


In NSTX H-mode plasmas, the confinement time is observed to be inversely proportional to the electron collisionality\cite{kaye2007confinement}, $\tau_E \frac{eB}{m_i}\sim \nu_*^{-(0.85-0.95)}$, with MTM-driven transport being a potential contributor to this scaling, notably in the weakly collisional regime. Contrary to this, our simulations of collisional MTMs show a different trend where higher collisionality stabilizes MTMs, while lower collisionality induces unstable MTMs, leading to increased thermal turbulence transport and decreased confinement. The different results in comparison with the experiment are due to the collision regime. Either very low or high collision stabilizes MTMs. Our model limit the investigation in the strong collisional regime. However, in the opposite regime, MTMs are unstable as collisonality increases, leading to a worse confinement. This finding encourages us to further investigate MTMs in the weakly collisional regime in our future research.

Hassam's model is valid in the strong collision regime, and it limits our simulation to reach out to the weak collision regime, where $\omega/\nu_{ei}\approx 1$. Unfortunately, in the edge pedestal of modern Tokamak devices, the electron diamagnetic frequency is close to the collision frequency, which means for the MTM $\omega/\nu_{ei}\approx1$. In the gyro-kinetic simulations of today's Tokamak devices, the most unstable mode is found to be $\omega/\nu_{ei}\approx 1$~\cite{curie2022gyrokinetic}. We also recognize additional mechanisms for MTM saturation, such as turbulence regulation by Zonal Flows (ZFs) and energy cascading due to nonlinear toroidal mode couplings\cite{wang2007nonlocal}. Our current BOUT++ simulations primarily focus on the relaxation of the electron temperature profile, partly because ZF calculations are still in development, resulting in their exclusion. Prioritizing the integration of ZFs in future research is essential to fully understand their regulatory role. We also plan to delve into the effects of energy cascading on MTM saturation, acknowledging its potentially significant impact. Our simulations are currently limited by the use of Hassam’s model, which assumes $\omega_{*e}/\nu_{ei} \ll 1$, confining our analysis to lower-n modes. Our objective is to advance a fluid model of MTM that can handle arbitrary frequency responses, enabling the inclusion of higher-n modes to extensively assess energy cascading.     


In our simulation, we are focused on investigating the unstable microtearing modes. To simplify the analysis, we have considered a limiting case with infinite $\eta_e$, thereby omitting the complexities associated with the density gradient and its stabilizing effects\cite{yagyu2023destabilization}. This method enables us to pinpoint the maximum growth rate, indicative of the most unstable limit. Moving forward, we plan to integrate the density gradient effects into our research to comprehensively understand their impact on MTM stability.

From the discussion above, we find that the electron thermal transport of tearing mode turbulence in fluid simulations is strongly influenced by the heat flux model/closure. Simulations with the Landau-fluid closure~\cite{chen2019extension, zhu2021drift} and perhaps more sophisticated parallel heat flux models~\cite{hunana2018new,wang2019landau,wang2020deep} may provide quantitatively more accurate calculation of electron thermal transport. Other nonlinear effects like energy cascading due to toroidal toroidal mode coupling are believed to relieve the issue. Coupling both ion modes and MTM in one simulation to investigate the thermal transport for both ions and electrons is planned as a future study. 

\section*{Acknowledgement}\label{sec:acknow}
The authors would like to thank Drs. N. Li, C. Xiao, Z. Guo, Y. Wang, C. Li, M. Curie for helpful discussions. K. Fan and C. Dong are supported by the National MCF Energy R\&D Program under Grant No. 2018YFE0311300. B. Zhu and X.Q. Xu are under the support by of the U.S. Department of Energy by Lawrence Livermore National Laboratory under Contract DE-AC52-07NA27344. T. Xia is supported by the National Natural Science Foundation of China 12175275, the Youth Innovation Promotion Association Chinese Academy of Sciences (Y2021114). This research used resources of the National Energy Research Scientific Computing Center, a DOE Office of Science User Facility supported by the Office of Science of the U.S. Department of Energy under Contract No. DE-AC02-05CH11231 using NERSC award FES-ERCAP0024033, FES-ERCAP0023237, FES-ERCAP0024539.

\section*{Data Availability Statement}
The data that support the findings of this study are available from the corresponding author upon reasonable request.

\section*{Appendix A: Laplacian equation solver}
\appendix{Laplacian equation solver}\label{sec:app1}
\setcounter{equation}{0}
\renewcommand{\theequation}{A\arabic{equation}}
As demonstrated in section \ref{sec:setup}, the equations having a form like Eq.(\ref{eq: invert lapalce psi}) are solved by the '\textit{invert-lapalce}' solver in BOUT++. This solver can solve not only the linear equations like Eq.(\ref{eq: invert lapalce psi}) and Eq.(\ref{eq: invert lapalce phi}), but also the nonlinear issue. In general, '\textit{invert-laplace}' solver is designed to solve an equation of the form,
\begin{equation}
    \mathbf{D} \nabla_\bot^2 f + \mathbf{A} f + \mathbf{b_0}\times \nabla f\cdot\nabla \mathbf{C} =\mathbf{B},
    \label{eq: laplacian eq}
\end{equation}
where \textbf{D}, \textbf{A}, and \textbf{C} are 2 dimension variables, e.g. $\textbf{D}=\textbf{D}(x,y)$ , and \textbf{B} is a 3 dimension variable, $\textbf{B}=\textbf{B}(x,y,z)$, $b_0$ is the direction of magnetic field line. The Eq.(\ref{eq: laplacian eq}) can be written in the form,
\begin{equation}
\begin{aligned}
        \textbf{D}\left(g^{xx}\frac{\partial^2}{\partial x^2} + G^x\frac{\partial}{\partial x} + g^{zz}\frac{\partial^2}{\partial z^2}+ G^z \frac{\partial}{\partial z }+ 2g^{xz}\frac{\partial^2}{\partial x\partial z}\right)f + \frac{\sqrt{g_{yy}}}{J}\frac{\partial }{\partial z}\textbf{C}\frac{\partial }{\partial x}f + \textbf{A}f=\textbf{B}.
        \label{eq: coef for matrix solver}
\end{aligned}
\end{equation}
Here, the difference parallel to the magnetic field line is neglected, $\partial/\partial y \approx  0$, $J$ is the Jacobi. Therefore, Eq.(\ref{eq: laplacian eq}) is reduced to a 2 dimension equation. After Fourier transforms along z,
\begin{equation}
    \begin{aligned}
        \textbf{D}\left(g^{xx}\frac{\partial^2}{\partial x^2} + G^x\frac{\partial}{\partial x} - g^{zz} k_z^2+ ik_z G^z + 2ik_zg^{xz}\frac{\partial}{\partial x}\right)f_{k_z} + ik_z\frac{\sqrt{g_{yy}}}{J}\textbf{C}\frac{\partial }{\partial x}f_{k_z} + \textbf{A}f_{k_z}=\textbf{B}_{k_z}.
        \label{eq: Laplacian eq after Fourier transform}
    \end{aligned}
\end{equation}
 After discrete along $x$ direction, Eq.(\ref{eq: Laplacian eq after Fourier transform}) is solved by a matrix method. In the nonlinear simulation of MTM, the coefficient \textbf{C} is the initial temperature $T_{e0}$ plus the zonal component of electron temperature  $\widetilde{T_e}|_{\text{DC}}$. 

\nocite{*}
\bibliography{MTM}

\begin{thebibliography}{44}%
\makeatletter
\providecommand \@ifxundefined [1]{%
 \@ifx{#1\undefined}
}%
\providecommand \@ifnum [1]{%
 \ifnum #1\expandafter \@firstoftwo
 \else \expandafter \@secondoftwo
 \fi
}%
\providecommand \@ifx [1]{%
 \ifx #1\expandafter \@firstoftwo
 \else \expandafter \@secondoftwo
 \fi
}%
\providecommand \natexlab [1]{#1}%
\providecommand \enquote  [1]{``#1''}%
\providecommand \bibnamefont  [1]{#1}%
\providecommand \bibfnamefont [1]{#1}%
\providecommand \citenamefont [1]{#1}%
\providecommand \href@noop [0]{\@secondoftwo}%
\providecommand \href [0]{\begingroup \@sanitize@url \@href}%
\providecommand \@href[1]{\@@startlink{#1}\@@href}%
\providecommand \@@href[1]{\endgroup#1\@@endlink}%
\providecommand \@sanitize@url [0]{\catcode `\\12\catcode `\$12\catcode
  `\&12\catcode `\#12\catcode `\^12\catcode `\_12\catcode `\%12\relax}%
\providecommand \@@startlink[1]{}%
\providecommand \@@endlink[0]{}%
\providecommand \url  [0]{\begingroup\@sanitize@url \@url }%
\providecommand \@url [1]{\endgroup\@href {#1}{\urlprefix }}%
\providecommand \urlprefix  [0]{URL }%
\providecommand \Eprint [0]{\href }%
\providecommand \doibase [0]{http://dx.doi.org/}%
\providecommand \selectlanguage [0]{\@gobble}%
\providecommand \bibinfo  [0]{\@secondoftwo}%
\providecommand \bibfield  [0]{\@secondoftwo}%
\providecommand \translation [1]{[#1]}%
\providecommand \BibitemOpen [0]{}%
\providecommand \bibitemStop [0]{}%
\providecommand \bibitemNoStop [0]{.\EOS\space}%
\providecommand \EOS [0]{\spacefactor3000\relax}%
\providecommand \BibitemShut  [1]{\csname bibitem#1\endcsname}%
\let\auto@bib@innerbib\@empty
\bibitem [{\citenamefont {Nelson}\ \emph {et~al.}(2021)\citenamefont {Nelson},
  \citenamefont {Laggner}, \citenamefont {Diallo}, \citenamefont {Smith},
  \citenamefont {Xing}, \citenamefont {Shousha},\ and\ \citenamefont
  {Kolemen}}]{nelson2021time}%
  \BibitemOpen
  \bibfield  {author} {\bibinfo {author} {\bibfnamefont {A.~O.}\ \bibnamefont
  {Nelson}}, \bibinfo {author} {\bibfnamefont {F.~M.}\ \bibnamefont {Laggner}},
  \bibinfo {author} {\bibfnamefont {A.}~\bibnamefont {Diallo}}, \bibinfo
  {author} {\bibfnamefont {D.}~\bibnamefont {Smith}}, \bibinfo {author}
  {\bibfnamefont {Z.~A.}\ \bibnamefont {Xing}}, \bibinfo {author}
  {\bibfnamefont {R.}~\bibnamefont {Shousha}}, \ and\ \bibinfo {author}
  {\bibfnamefont {E.}~\bibnamefont {Kolemen}},\ }\href@noop {} {\bibfield
  {journal} {\bibinfo  {journal} {Nuclear Fusion}\ }\textbf {\bibinfo {volume}
  {61}},\ \bibinfo {pages} {116038} (\bibinfo {year} {2021})}\BibitemShut
  {NoStop}%
\bibitem [{\citenamefont {Hatch}\ \emph {et~al.}(2021)\citenamefont {Hatch},
  \citenamefont {Kotschenreuther}, \citenamefont {Mahajan}, \citenamefont
  {Pueschel}, \citenamefont {Michoski}, \citenamefont {Merlo}, \citenamefont
  {Hassan}, \citenamefont {Field}, \citenamefont {Frassinetti}, \citenamefont
  {Giroud} \emph {et~al.}}]{hatch2021microtearing}%
  \BibitemOpen
  \bibfield  {author} {\bibinfo {author} {\bibfnamefont {D.}~\bibnamefont
  {Hatch}}, \bibinfo {author} {\bibfnamefont {M.}~\bibnamefont
  {Kotschenreuther}}, \bibinfo {author} {\bibfnamefont {S.}~\bibnamefont
  {Mahajan}}, \bibinfo {author} {\bibfnamefont {M.}~\bibnamefont {Pueschel}},
  \bibinfo {author} {\bibfnamefont {C.}~\bibnamefont {Michoski}}, \bibinfo
  {author} {\bibfnamefont {G.}~\bibnamefont {Merlo}}, \bibinfo {author}
  {\bibfnamefont {E.}~\bibnamefont {Hassan}}, \bibinfo {author} {\bibfnamefont
  {A.}~\bibnamefont {Field}}, \bibinfo {author} {\bibfnamefont
  {L.}~\bibnamefont {Frassinetti}}, \bibinfo {author} {\bibfnamefont
  {C.}~\bibnamefont {Giroud}},  \emph {et~al.},\ }\href@noop {} {\bibfield
  {journal} {\bibinfo  {journal} {Nuclear Fusion}\ }\textbf {\bibinfo {volume}
  {61}},\ \bibinfo {pages} {036015} (\bibinfo {year} {2021})}\BibitemShut
  {NoStop}%
\bibitem [{\citenamefont {Hatch}\ \emph {et~al.}(2016)\citenamefont {Hatch},
  \citenamefont {Kotschenreuther}, \citenamefont {Mahajan}, \citenamefont
  {Valanju}, \citenamefont {Jenko}, \citenamefont {Told}, \citenamefont
  {G{\"o}rler},\ and\ \citenamefont {Saarelma}}]{hatch2016microtearing}%
  \BibitemOpen
  \bibfield  {author} {\bibinfo {author} {\bibfnamefont {D.}~\bibnamefont
  {Hatch}}, \bibinfo {author} {\bibfnamefont {M.}~\bibnamefont
  {Kotschenreuther}}, \bibinfo {author} {\bibfnamefont {S.}~\bibnamefont
  {Mahajan}}, \bibinfo {author} {\bibfnamefont {P.}~\bibnamefont {Valanju}},
  \bibinfo {author} {\bibfnamefont {F.}~\bibnamefont {Jenko}}, \bibinfo
  {author} {\bibfnamefont {D.}~\bibnamefont {Told}}, \bibinfo {author}
  {\bibfnamefont {T.}~\bibnamefont {G{\"o}rler}}, \ and\ \bibinfo {author}
  {\bibfnamefont {S.}~\bibnamefont {Saarelma}},\ }\href@noop {} {\bibfield
  {journal} {\bibinfo  {journal} {Nuclear Fusion}\ }\textbf {\bibinfo {volume}
  {56}},\ \bibinfo {pages} {104003} (\bibinfo {year} {2016})}\BibitemShut
  {NoStop}%
\bibitem [{\citenamefont {Chen}\ \emph {et~al.}(2023)\citenamefont {Chen},
  \citenamefont {Jian}, \citenamefont {Brower}, \citenamefont {Haskey},
  \citenamefont {Yan}, \citenamefont {Groebner}, \citenamefont {Wang},
  \citenamefont {Rhodes}, \citenamefont {Laggner}, \citenamefont {Ding} \emph
  {et~al.}}]{chen2023micro}%
  \BibitemOpen
  \bibfield  {author} {\bibinfo {author} {\bibfnamefont {J.}~\bibnamefont
  {Chen}}, \bibinfo {author} {\bibfnamefont {X.}~\bibnamefont {Jian}}, \bibinfo
  {author} {\bibfnamefont {D.}~\bibnamefont {Brower}}, \bibinfo {author}
  {\bibfnamefont {S.}~\bibnamefont {Haskey}}, \bibinfo {author} {\bibfnamefont
  {Z.}~\bibnamefont {Yan}}, \bibinfo {author} {\bibfnamefont {R.}~\bibnamefont
  {Groebner}}, \bibinfo {author} {\bibfnamefont {H.}~\bibnamefont {Wang}},
  \bibinfo {author} {\bibfnamefont {T.}~\bibnamefont {Rhodes}}, \bibinfo
  {author} {\bibfnamefont {F.}~\bibnamefont {Laggner}}, \bibinfo {author}
  {\bibfnamefont {W.}~\bibnamefont {Ding}},  \emph {et~al.},\ }\href@noop {}
  {\bibfield  {journal} {\bibinfo  {journal} {Nuclear Fusion}\ }\textbf
  {\bibinfo {volume} {63}},\ \bibinfo {pages} {066019} (\bibinfo {year}
  {2023})}\BibitemShut {NoStop}%
\bibitem [{\citenamefont {Jian}\ \emph {et~al.}(2019)\citenamefont {Jian},
  \citenamefont {Holland}, \citenamefont {Candy}, \citenamefont {Belli},
  \citenamefont {Chan}, \citenamefont {Garofalo},\ and\ \citenamefont
  {Ding}}]{jian2019role}%
  \BibitemOpen
  \bibfield  {author} {\bibinfo {author} {\bibfnamefont {X.}~\bibnamefont
  {Jian}}, \bibinfo {author} {\bibfnamefont {C.}~\bibnamefont {Holland}},
  \bibinfo {author} {\bibfnamefont {J.}~\bibnamefont {Candy}}, \bibinfo
  {author} {\bibfnamefont {E.}~\bibnamefont {Belli}}, \bibinfo {author}
  {\bibfnamefont {V.}~\bibnamefont {Chan}}, \bibinfo {author} {\bibfnamefont
  {A.~M.}\ \bibnamefont {Garofalo}}, \ and\ \bibinfo {author} {\bibfnamefont
  {S.}~\bibnamefont {Ding}},\ }\href@noop {} {\bibfield  {journal} {\bibinfo
  {journal} {Physical review letters}\ }\textbf {\bibinfo {volume} {123}},\
  \bibinfo {pages} {225002} (\bibinfo {year} {2019})}\BibitemShut {NoStop}%
\bibitem [{\citenamefont {Kaye}\ \emph {et~al.}(2014)\citenamefont {Kaye},
  \citenamefont {Guttenfelder}, \citenamefont {Bell}, \citenamefont {Gerhardt},
  \citenamefont {LeBlanc},\ and\ \citenamefont {Maingi}}]{kaye2014reduced}%
  \BibitemOpen
  \bibfield  {author} {\bibinfo {author} {\bibfnamefont {S.}~\bibnamefont
  {Kaye}}, \bibinfo {author} {\bibfnamefont {W.}~\bibnamefont {Guttenfelder}},
  \bibinfo {author} {\bibfnamefont {R.}~\bibnamefont {Bell}}, \bibinfo {author}
  {\bibfnamefont {S.}~\bibnamefont {Gerhardt}}, \bibinfo {author}
  {\bibfnamefont {B.}~\bibnamefont {LeBlanc}}, \ and\ \bibinfo {author}
  {\bibfnamefont {R.}~\bibnamefont {Maingi}},\ }\href@noop {} {\bibfield
  {journal} {\bibinfo  {journal} {Physics of Plasmas}\ }\textbf {\bibinfo
  {volume} {21}} (\bibinfo {year} {2014})}\BibitemShut {NoStop}%
\bibitem [{\citenamefont {Guttenfelder}\ \emph {et~al.}(2011)\citenamefont
  {Guttenfelder}, \citenamefont {Candy}, \citenamefont {Kaye}, \citenamefont
  {Nevins}, \citenamefont {Wang}, \citenamefont {Bell}, \citenamefont
  {Hammett}, \citenamefont {LeBlanc}, \citenamefont {Mikkelsen},\ and\
  \citenamefont {Yuh}}]{guttenfelder2011electromagnetic}%
  \BibitemOpen
  \bibfield  {author} {\bibinfo {author} {\bibfnamefont {W.}~\bibnamefont
  {Guttenfelder}}, \bibinfo {author} {\bibfnamefont {J.}~\bibnamefont {Candy}},
  \bibinfo {author} {\bibfnamefont {S.}~\bibnamefont {Kaye}}, \bibinfo {author}
  {\bibfnamefont {W.}~\bibnamefont {Nevins}}, \bibinfo {author} {\bibfnamefont
  {E.}~\bibnamefont {Wang}}, \bibinfo {author} {\bibfnamefont {R.}~\bibnamefont
  {Bell}}, \bibinfo {author} {\bibfnamefont {G.}~\bibnamefont {Hammett}},
  \bibinfo {author} {\bibfnamefont {B.}~\bibnamefont {LeBlanc}}, \bibinfo
  {author} {\bibfnamefont {D.}~\bibnamefont {Mikkelsen}}, \ and\ \bibinfo
  {author} {\bibfnamefont {H.}~\bibnamefont {Yuh}},\ }\href@noop {} {\bibfield
  {journal} {\bibinfo  {journal} {Physical review letters}\ }\textbf {\bibinfo
  {volume} {106}},\ \bibinfo {pages} {155004} (\bibinfo {year}
  {2011})}\BibitemShut {NoStop}%
\bibitem [{\citenamefont {Hazeltine}\ \emph {et~al.}(1975)\citenamefont
  {Hazeltine}, \citenamefont {Dobrott},\ and\ \citenamefont
  {Wang}}]{hazeltine1975kinetic}%
  \BibitemOpen
  \bibfield  {author} {\bibinfo {author} {\bibfnamefont {R.}~\bibnamefont
  {Hazeltine}}, \bibinfo {author} {\bibfnamefont {D.}~\bibnamefont {Dobrott}},
  \ and\ \bibinfo {author} {\bibfnamefont {T.}~\bibnamefont {Wang}},\
  }\href@noop {} {\bibfield  {journal} {\bibinfo  {journal} {The Physics of
  Fluids}\ }\textbf {\bibinfo {volume} {18}},\ \bibinfo {pages} {1778}
  (\bibinfo {year} {1975})}\BibitemShut {NoStop}%
\bibitem [{\citenamefont {Drake}\ and\ \citenamefont
  {Lee}(1977)}]{drake1977kinetic}%
  \BibitemOpen
  \bibfield  {author} {\bibinfo {author} {\bibfnamefont {J.}~\bibnamefont
  {Drake}}\ and\ \bibinfo {author} {\bibfnamefont {Y.}~\bibnamefont {Lee}},\
  }\href@noop {} {\bibfield  {journal} {\bibinfo  {journal} {The Physics of
  Fluids}\ }\textbf {\bibinfo {volume} {20}},\ \bibinfo {pages} {1341}
  (\bibinfo {year} {1977})}\BibitemShut {NoStop}%
\bibitem [{\citenamefont {Larakers}\ \emph {et~al.}(2020)\citenamefont
  {Larakers}, \citenamefont {Hazeltine},\ and\ \citenamefont
  {Mahajan}}]{larakers2020comprehensive}%
  \BibitemOpen
  \bibfield  {author} {\bibinfo {author} {\bibfnamefont {J.}~\bibnamefont
  {Larakers}}, \bibinfo {author} {\bibfnamefont {R.}~\bibnamefont {Hazeltine}},
  \ and\ \bibinfo {author} {\bibfnamefont {S.}~\bibnamefont {Mahajan}},\
  }\href@noop {} {\bibfield  {journal} {\bibinfo  {journal} {Physics of
  Plasmas}\ }\textbf {\bibinfo {volume} {27}} (\bibinfo {year}
  {2020})}\BibitemShut {NoStop}%
\bibitem [{\citenamefont {Larakers}\ \emph {et~al.}(2021)\citenamefont
  {Larakers}, \citenamefont {Curie}, \citenamefont {Hatch}, \citenamefont
  {Hazeltine},\ and\ \citenamefont {Mahajan}}]{larakers2021global}%
  \BibitemOpen
  \bibfield  {author} {\bibinfo {author} {\bibfnamefont {J.}~\bibnamefont
  {Larakers}}, \bibinfo {author} {\bibfnamefont {M.}~\bibnamefont {Curie}},
  \bibinfo {author} {\bibfnamefont {D.}~\bibnamefont {Hatch}}, \bibinfo
  {author} {\bibfnamefont {R.}~\bibnamefont {Hazeltine}}, \ and\ \bibinfo
  {author} {\bibfnamefont {S.}~\bibnamefont {Mahajan}},\ }\href@noop {}
  {\bibfield  {journal} {\bibinfo  {journal} {Physical Review Letters}\
  }\textbf {\bibinfo {volume} {126}},\ \bibinfo {pages} {225001} (\bibinfo
  {year} {2021})}\BibitemShut {NoStop}%
\bibitem [{\citenamefont {Dudson}\ \emph {et~al.}(2009)\citenamefont {Dudson},
  \citenamefont {Umansky}, \citenamefont {Xu}, \citenamefont {Snyder},\ and\
  \citenamefont {Wilson}}]{dudson2009bout++}%
  \BibitemOpen
  \bibfield  {author} {\bibinfo {author} {\bibfnamefont {B.}~\bibnamefont
  {Dudson}}, \bibinfo {author} {\bibfnamefont {M.}~\bibnamefont {Umansky}},
  \bibinfo {author} {\bibfnamefont {X.}~\bibnamefont {Xu}}, \bibinfo {author}
  {\bibfnamefont {P.}~\bibnamefont {Snyder}}, \ and\ \bibinfo {author}
  {\bibfnamefont {H.}~\bibnamefont {Wilson}},\ }\href@noop {} {\bibfield
  {journal} {\bibinfo  {journal} {Computer Physics Communications}\ }\textbf
  {\bibinfo {volume} {180}},\ \bibinfo {pages} {1467} (\bibinfo {year}
  {2009})}\BibitemShut {NoStop}%
\bibitem [{\citenamefont {Xi}\ \emph {et~al.}(2014)\citenamefont {Xi},
  \citenamefont {Xu},\ and\ \citenamefont {Diamond}}]{xi2014phase}%
  \BibitemOpen
  \bibfield  {author} {\bibinfo {author} {\bibfnamefont {P.}~\bibnamefont
  {Xi}}, \bibinfo {author} {\bibfnamefont {X.}~\bibnamefont {Xu}}, \ and\
  \bibinfo {author} {\bibfnamefont {P.}~\bibnamefont {Diamond}},\ }\href@noop
  {} {\bibfield  {journal} {\bibinfo  {journal} {Physical Review Letters}\
  }\textbf {\bibinfo {volume} {112}},\ \bibinfo {pages} {085001} (\bibinfo
  {year} {2014})}\BibitemShut {NoStop}%
\bibitem [{\citenamefont {Xu}\ \emph {et~al.}(2019)\citenamefont {Xu},
  \citenamefont {Li}, \citenamefont {Li}, \citenamefont {Chen}, \citenamefont
  {Xia}, \citenamefont {Tang}, \citenamefont {Zhu},\ and\ \citenamefont
  {Chan}}]{xu2019simulations}%
  \BibitemOpen
  \bibfield  {author} {\bibinfo {author} {\bibfnamefont {X.}~\bibnamefont
  {Xu}}, \bibinfo {author} {\bibfnamefont {N.}~\bibnamefont {Li}}, \bibinfo
  {author} {\bibfnamefont {Z.}~\bibnamefont {Li}}, \bibinfo {author}
  {\bibfnamefont {B.}~\bibnamefont {Chen}}, \bibinfo {author} {\bibfnamefont
  {T.}~\bibnamefont {Xia}}, \bibinfo {author} {\bibfnamefont {T.}~\bibnamefont
  {Tang}}, \bibinfo {author} {\bibfnamefont {B.}~\bibnamefont {Zhu}}, \ and\
  \bibinfo {author} {\bibfnamefont {V.}~\bibnamefont {Chan}},\ }\href@noop {}
  {\bibfield  {journal} {\bibinfo  {journal} {Nuclear Fusion}\ }\textbf
  {\bibinfo {volume} {59}},\ \bibinfo {pages} {126039} (\bibinfo {year}
  {2019})}\BibitemShut {NoStop}%
\bibitem [{\citenamefont {Hassam}(1980{\natexlab{a}})}]{hassam1980higher}%
  \BibitemOpen
  \bibfield  {author} {\bibinfo {author} {\bibfnamefont {A.}~\bibnamefont
  {Hassam}},\ }\href@noop {} {\bibfield  {journal} {\bibinfo  {journal} {The
  Physics of Fluids}\ }\textbf {\bibinfo {volume} {23}},\ \bibinfo {pages} {38}
  (\bibinfo {year} {1980}{\natexlab{a}})}\BibitemShut {NoStop}%
\bibitem [{\citenamefont {Drake}\ \emph {et~al.}(1980)\citenamefont {Drake},
  \citenamefont {Gladd}, \citenamefont {Liu},\ and\ \citenamefont
  {Chang}}]{drake1980microtearing}%
  \BibitemOpen
  \bibfield  {author} {\bibinfo {author} {\bibfnamefont {J.}~\bibnamefont
  {Drake}}, \bibinfo {author} {\bibfnamefont {N.}~\bibnamefont {Gladd}},
  \bibinfo {author} {\bibfnamefont {C.}~\bibnamefont {Liu}}, \ and\ \bibinfo
  {author} {\bibfnamefont {C.}~\bibnamefont {Chang}},\ }\href@noop {}
  {\bibfield  {journal} {\bibinfo  {journal} {Physical review letters}\
  }\textbf {\bibinfo {volume} {44}},\ \bibinfo {pages} {994} (\bibinfo {year}
  {1980})}\BibitemShut {NoStop}%
\bibitem [{\citenamefont {Hassam}(1980{\natexlab{b}})}]{hassam1980fluid}%
  \BibitemOpen
  \bibfield  {author} {\bibinfo {author} {\bibfnamefont {A.}~\bibnamefont
  {Hassam}},\ }\href@noop {} {\bibfield  {journal} {\bibinfo  {journal} {The
  Physics of Fluids}\ }\textbf {\bibinfo {volume} {23}},\ \bibinfo {pages}
  {2493} (\bibinfo {year} {1980}{\natexlab{b}})}\BibitemShut {NoStop}%
\bibitem [{\citenamefont {Gladd}\ \emph {et~al.}(1980)\citenamefont {Gladd},
  \citenamefont {Drake}, \citenamefont {Chang},\ and\ \citenamefont
  {Liu}}]{gladd1980electron}%
  \BibitemOpen
  \bibfield  {author} {\bibinfo {author} {\bibfnamefont {N.}~\bibnamefont
  {Gladd}}, \bibinfo {author} {\bibfnamefont {J.}~\bibnamefont {Drake}},
  \bibinfo {author} {\bibfnamefont {C.}~\bibnamefont {Chang}}, \ and\ \bibinfo
  {author} {\bibfnamefont {C.}~\bibnamefont {Liu}},\ }\href@noop {} {\bibfield
  {journal} {\bibinfo  {journal} {The Physics of Fluids}\ }\textbf {\bibinfo
  {volume} {23}},\ \bibinfo {pages} {1182} (\bibinfo {year}
  {1980})}\BibitemShut {NoStop}%
\bibitem [{\citenamefont {D’Ippolito}\ \emph {et~al.}(1980)\citenamefont
  {D’Ippolito}, \citenamefont {Lee},\ and\ \citenamefont
  {Drake}}]{d1980linear}%
  \BibitemOpen
  \bibfield  {author} {\bibinfo {author} {\bibfnamefont {D.}~\bibnamefont
  {D’Ippolito}}, \bibinfo {author} {\bibfnamefont {Y.}~\bibnamefont {Lee}}, \
  and\ \bibinfo {author} {\bibfnamefont {J.}~\bibnamefont {Drake}},\
  }\href@noop {} {\bibfield  {journal} {\bibinfo  {journal} {The Physics of
  Fluids}\ }\textbf {\bibinfo {volume} {23}},\ \bibinfo {pages} {771} (\bibinfo
  {year} {1980})}\BibitemShut {NoStop}%
\bibitem [{\citenamefont {Furth}\ \emph {et~al.}(1963)\citenamefont {Furth},
  \citenamefont {Killeen},\ and\ \citenamefont {Rosenbluth}}]{furth1963finite}%
  \BibitemOpen
  \bibfield  {author} {\bibinfo {author} {\bibfnamefont {H.~P.}\ \bibnamefont
  {Furth}}, \bibinfo {author} {\bibfnamefont {J.}~\bibnamefont {Killeen}}, \
  and\ \bibinfo {author} {\bibfnamefont {M.~N.}\ \bibnamefont {Rosenbluth}},\
  }\href@noop {} {\bibfield  {journal} {\bibinfo  {journal} {The physics of
  Fluids}\ }\textbf {\bibinfo {volume} {6}},\ \bibinfo {pages} {459} (\bibinfo
  {year} {1963})}\BibitemShut {NoStop}%
\bibitem [{\citenamefont {Zhu}\ \emph {et~al.}(2021)\citenamefont {Zhu},
  \citenamefont {Seto}, \citenamefont {Xu},\ and\ \citenamefont
  {Yagi}}]{zhu2021drift}%
  \BibitemOpen
  \bibfield  {author} {\bibinfo {author} {\bibfnamefont {B.}~\bibnamefont
  {Zhu}}, \bibinfo {author} {\bibfnamefont {H.}~\bibnamefont {Seto}}, \bibinfo
  {author} {\bibfnamefont {X.-q.}\ \bibnamefont {Xu}}, \ and\ \bibinfo {author}
  {\bibfnamefont {M.}~\bibnamefont {Yagi}},\ }\href@noop {} {\bibfield
  {journal} {\bibinfo  {journal} {Computer Physics Communications}\ }\textbf
  {\bibinfo {volume} {267}},\ \bibinfo {pages} {108079} (\bibinfo {year}
  {2021})}\BibitemShut {NoStop}%
\bibitem [{\citenamefont {Braginskii}(1965)}]{braginskii1965transport}%
  \BibitemOpen
  \bibfield  {author} {\bibinfo {author} {\bibfnamefont {S.}~\bibnamefont
  {Braginskii}},\ }\href@noop {} {\bibfield  {journal} {\bibinfo  {journal}
  {Reviews of plasma physics}\ }\textbf {\bibinfo {volume} {1}},\ \bibinfo
  {pages} {205} (\bibinfo {year} {1965})}\BibitemShut {NoStop}%
\bibitem [{\citenamefont {Zhu}\ \emph {et~al.}(2023)\citenamefont {Zhu},
  \citenamefont {Xu},\ and\ \citenamefont {Tang}}]{zhu2023electromagnetic}%
  \BibitemOpen
  \bibfield  {author} {\bibinfo {author} {\bibfnamefont {B.}~\bibnamefont
  {Zhu}}, \bibinfo {author} {\bibfnamefont {X.}~\bibnamefont {Xu}}, \ and\
  \bibinfo {author} {\bibfnamefont {X.}~\bibnamefont {Tang}},\ }\href@noop {}
  {\bibfield  {journal} {\bibinfo  {journal} {Nuclear Fusion}\ } (\bibinfo
  {year} {2023})}\BibitemShut {NoStop}%
\bibitem [{\citenamefont {Xu}\ \emph {et~al.}(2011)\citenamefont {Xu},
  \citenamefont {Dudson}, \citenamefont {Snyder}, \citenamefont {Umansky},
  \citenamefont {Wilson},\ and\ \citenamefont {Casper}}]{xu2011nonlinear}%
  \BibitemOpen
  \bibfield  {author} {\bibinfo {author} {\bibfnamefont {X.}~\bibnamefont
  {Xu}}, \bibinfo {author} {\bibfnamefont {B.}~\bibnamefont {Dudson}}, \bibinfo
  {author} {\bibfnamefont {P.}~\bibnamefont {Snyder}}, \bibinfo {author}
  {\bibfnamefont {M.}~\bibnamefont {Umansky}}, \bibinfo {author} {\bibfnamefont
  {H.}~\bibnamefont {Wilson}}, \ and\ \bibinfo {author} {\bibfnamefont
  {T.}~\bibnamefont {Casper}},\ }\href@noop {} {\bibfield  {journal} {\bibinfo
  {journal} {Nuclear Fusion}\ }\textbf {\bibinfo {volume} {51}},\ \bibinfo
  {pages} {103040} (\bibinfo {year} {2011})}\BibitemShut {NoStop}%
\bibitem [{\citenamefont {Xu}\ \emph {et~al.}(2010)\citenamefont {Xu},
  \citenamefont {Dudson}, \citenamefont {Snyder}, \citenamefont {Umansky},\
  and\ \citenamefont {Wilson}}]{xu2010nonlinear}%
  \BibitemOpen
  \bibfield  {author} {\bibinfo {author} {\bibfnamefont {X.}~\bibnamefont
  {Xu}}, \bibinfo {author} {\bibfnamefont {B.}~\bibnamefont {Dudson}}, \bibinfo
  {author} {\bibfnamefont {P.}~\bibnamefont {Snyder}}, \bibinfo {author}
  {\bibfnamefont {M.}~\bibnamefont {Umansky}}, \ and\ \bibinfo {author}
  {\bibfnamefont {H.}~\bibnamefont {Wilson}},\ }\href@noop {} {\bibfield
  {journal} {\bibinfo  {journal} {Physical review letters}\ }\textbf {\bibinfo
  {volume} {105}},\ \bibinfo {pages} {175005} (\bibinfo {year}
  {2010})}\BibitemShut {NoStop}%
\bibitem [{\citenamefont {Zhu}\ \emph {et~al.}(2018)\citenamefont {Zhu},
  \citenamefont {Francisquez},\ and\ \citenamefont {Rogers}}]{zhu2018up}%
  \BibitemOpen
  \bibfield  {author} {\bibinfo {author} {\bibfnamefont {B.}~\bibnamefont
  {Zhu}}, \bibinfo {author} {\bibfnamefont {M.}~\bibnamefont {Francisquez}}, \
  and\ \bibinfo {author} {\bibfnamefont {B.~N.}\ \bibnamefont {Rogers}},\
  }\href@noop {} {\bibfield  {journal} {\bibinfo  {journal} {Nuclear Fusion}\
  }\textbf {\bibinfo {volume} {58}},\ \bibinfo {pages} {106039} (\bibinfo
  {year} {2018})}\BibitemShut {NoStop}%
\bibitem [{\citenamefont {Zhu}\ \emph {et~al.}(2017)\citenamefont {Zhu},
  \citenamefont {Francisquez},\ and\ \citenamefont {Rogers}}]{zhu2017global}%
  \BibitemOpen
  \bibfield  {author} {\bibinfo {author} {\bibfnamefont {B.}~\bibnamefont
  {Zhu}}, \bibinfo {author} {\bibfnamefont {M.}~\bibnamefont {Francisquez}}, \
  and\ \bibinfo {author} {\bibfnamefont {B.~N.}\ \bibnamefont {Rogers}},\
  }\href@noop {} {\bibfield  {journal} {\bibinfo  {journal} {Physics of
  Plasmas}\ }\textbf {\bibinfo {volume} {24}} (\bibinfo {year}
  {2017})}\BibitemShut {NoStop}%
\bibitem [{\citenamefont {Crotinger}\ \emph {et~al.}(1997)\citenamefont
  {Crotinger}, \citenamefont {LoDestro}, \citenamefont {Pearlstein},
  \citenamefont {Tarditi}, \citenamefont {Casper},\ and\ \citenamefont
  {Hooper}}]{crotinger1997corsica}%
  \BibitemOpen
  \bibfield  {author} {\bibinfo {author} {\bibfnamefont {J.~A.}\ \bibnamefont
  {Crotinger}}, \bibinfo {author} {\bibfnamefont {L.}~\bibnamefont {LoDestro}},
  \bibinfo {author} {\bibfnamefont {L.~D.}\ \bibnamefont {Pearlstein}},
  \bibinfo {author} {\bibfnamefont {A.}~\bibnamefont {Tarditi}}, \bibinfo
  {author} {\bibfnamefont {T.}~\bibnamefont {Casper}}, \ and\ \bibinfo {author}
  {\bibfnamefont {E.~B.}\ \bibnamefont {Hooper}},\ }\href@noop {} {\emph
  {\bibinfo {title} {Corsica: A comprehensive simulation of toroidal
  magnetic-fusion devices. final report to the ldrd program}}},\ \bibinfo
  {type} {Tech. Rep.}\ (\bibinfo  {institution} {Lawrence Livermore National
  Lab.(LLNL), Livermore, CA (United States)},\ \bibinfo {year}
  {1997})\BibitemShut {NoStop}%
\bibitem [{\citenamefont {Seto}\ \emph {et~al.}(2023)\citenamefont {Seto},
  \citenamefont {Dudson}, \citenamefont {Xu},\ and\ \citenamefont
  {Yagi}}]{seto2023bout++}%
  \BibitemOpen
  \bibfield  {author} {\bibinfo {author} {\bibfnamefont {H.}~\bibnamefont
  {Seto}}, \bibinfo {author} {\bibfnamefont {B.~D.}\ \bibnamefont {Dudson}},
  \bibinfo {author} {\bibfnamefont {X.-Q.}\ \bibnamefont {Xu}}, \ and\ \bibinfo
  {author} {\bibfnamefont {M.}~\bibnamefont {Yagi}},\ }\href@noop {} {\bibfield
   {journal} {\bibinfo  {journal} {Computer Physics Communications}\ }\textbf
  {\bibinfo {volume} {283}},\ \bibinfo {pages} {108568} (\bibinfo {year}
  {2023})}\BibitemShut {NoStop}%
\bibitem [{\citenamefont {Xu}\ \emph {et~al.}(2008)\citenamefont {Xu},
  \citenamefont {Umansky}, \citenamefont {Dudson},\ and\ \citenamefont
  {Snyder}}]{xu2008boundary}%
  \BibitemOpen
  \bibfield  {author} {\bibinfo {author} {\bibfnamefont {X.}~\bibnamefont
  {Xu}}, \bibinfo {author} {\bibfnamefont {M.}~\bibnamefont {Umansky}},
  \bibinfo {author} {\bibfnamefont {B.}~\bibnamefont {Dudson}}, \ and\ \bibinfo
  {author} {\bibfnamefont {P.}~\bibnamefont {Snyder}},\ }\href@noop {}
  {\bibfield  {journal} {\bibinfo  {journal} {Communications in Computational
  Physics, vol. 4, no. 5, July 1, 2008, pp. 949-979}\ }\textbf {\bibinfo
  {volume} {4}} (\bibinfo {year} {2008})}\BibitemShut {NoStop}%
\bibitem [{\citenamefont {Yagyu}\ and\ \citenamefont
  {Numata}(2023)}]{yagyu2023destabilization}%
  \BibitemOpen
  \bibfield  {author} {\bibinfo {author} {\bibfnamefont {M.}~\bibnamefont
  {Yagyu}}\ and\ \bibinfo {author} {\bibfnamefont {R.}~\bibnamefont {Numata}},\
  }\href@noop {} {\bibfield  {journal} {\bibinfo  {journal} {Plasma Physics and
  Controlled Fusion}\ }\textbf {\bibinfo {volume} {65}},\ \bibinfo {pages}
  {065003} (\bibinfo {year} {2023})}\BibitemShut {NoStop}%
\bibitem [{\citenamefont {Curie}\ \emph
  {et~al.}(2022{\natexlab{a}})\citenamefont {Curie}, \citenamefont {Larakers},
  \citenamefont {Hatch}, \citenamefont {Nelson}, \citenamefont {Diallo},
  \citenamefont {Hassan}, \citenamefont {Guttenfelder}, \citenamefont
  {Halfmoon}, \citenamefont {Kotschenreuther}, \citenamefont {Hazeltine} \emph
  {et~al.}}]{curie2022survey}%
  \BibitemOpen
  \bibfield  {author} {\bibinfo {author} {\bibfnamefont {M.}~\bibnamefont
  {Curie}}, \bibinfo {author} {\bibfnamefont {J.}~\bibnamefont {Larakers}},
  \bibinfo {author} {\bibfnamefont {D.}~\bibnamefont {Hatch}}, \bibinfo
  {author} {\bibfnamefont {A.}~\bibnamefont {Nelson}}, \bibinfo {author}
  {\bibfnamefont {A.}~\bibnamefont {Diallo}}, \bibinfo {author} {\bibfnamefont
  {E.}~\bibnamefont {Hassan}}, \bibinfo {author} {\bibfnamefont
  {W.}~\bibnamefont {Guttenfelder}}, \bibinfo {author} {\bibfnamefont
  {M.}~\bibnamefont {Halfmoon}}, \bibinfo {author} {\bibfnamefont
  {M.}~\bibnamefont {Kotschenreuther}}, \bibinfo {author} {\bibfnamefont
  {R.}~\bibnamefont {Hazeltine}},  \emph {et~al.},\ }\href@noop {} {\bibfield
  {journal} {\bibinfo  {journal} {Physics of Plasmas}\ }\textbf {\bibinfo
  {volume} {29}} (\bibinfo {year} {2022}{\natexlab{a}})}\BibitemShut {NoStop}%
\bibitem [{\citenamefont {Curie}\ \emph {et~al.}(2023)\citenamefont {Curie},
  \citenamefont {Larakers}, \citenamefont {Parisi}, \citenamefont {Staebler},
  \citenamefont {Munaretto}, \citenamefont {Guttenfelder}, \citenamefont
  {Belli}, \citenamefont {Hatch}, \citenamefont {Lampert}, \citenamefont
  {Avdeeva} \emph {et~al.}}]{curie2023microtearding}%
  \BibitemOpen
  \bibfield  {author} {\bibinfo {author} {\bibfnamefont {M.~T.}\ \bibnamefont
  {Curie}}, \bibinfo {author} {\bibfnamefont {J.}~\bibnamefont {Larakers}},
  \bibinfo {author} {\bibfnamefont {J.}~\bibnamefont {Parisi}}, \bibinfo
  {author} {\bibfnamefont {G.}~\bibnamefont {Staebler}}, \bibinfo {author}
  {\bibfnamefont {S.}~\bibnamefont {Munaretto}}, \bibinfo {author}
  {\bibfnamefont {W.}~\bibnamefont {Guttenfelder}}, \bibinfo {author}
  {\bibfnamefont {E.}~\bibnamefont {Belli}}, \bibinfo {author} {\bibfnamefont
  {D.~R.}\ \bibnamefont {Hatch}}, \bibinfo {author} {\bibfnamefont
  {M.}~\bibnamefont {Lampert}}, \bibinfo {author} {\bibfnamefont
  {G.}~\bibnamefont {Avdeeva}},  \emph {et~al.},\ }\href@noop {} {\bibfield
  {journal} {\bibinfo  {journal} {arXiv preprint arXiv:2304.08982}\ } (\bibinfo
  {year} {2023})}\BibitemShut {NoStop}%
\bibitem [{\citenamefont {Hassan}\ \emph {et~al.}(2021)\citenamefont {Hassan},
  \citenamefont {Hatch}, \citenamefont {Halfmoon}, \citenamefont {Curie},
  \citenamefont {Kotchenreuther}, \citenamefont {Mahajan}, \citenamefont
  {Merlo}, \citenamefont {Groebner}, \citenamefont {Nelson},\ and\
  \citenamefont {Diallo}}]{hassan2021identifying}%
  \BibitemOpen
  \bibfield  {author} {\bibinfo {author} {\bibfnamefont {E.}~\bibnamefont
  {Hassan}}, \bibinfo {author} {\bibfnamefont {D.}~\bibnamefont {Hatch}},
  \bibinfo {author} {\bibfnamefont {M.}~\bibnamefont {Halfmoon}}, \bibinfo
  {author} {\bibfnamefont {M.}~\bibnamefont {Curie}}, \bibinfo {author}
  {\bibfnamefont {M.}~\bibnamefont {Kotchenreuther}}, \bibinfo {author}
  {\bibfnamefont {S.}~\bibnamefont {Mahajan}}, \bibinfo {author} {\bibfnamefont
  {G.}~\bibnamefont {Merlo}}, \bibinfo {author} {\bibfnamefont
  {R.}~\bibnamefont {Groebner}}, \bibinfo {author} {\bibfnamefont
  {A.}~\bibnamefont {Nelson}}, \ and\ \bibinfo {author} {\bibfnamefont
  {A.}~\bibnamefont {Diallo}},\ }\href@noop {} {\bibfield  {journal} {\bibinfo
  {journal} {Nuclear Fusion}\ }\textbf {\bibinfo {volume} {62}},\ \bibinfo
  {pages} {026008} (\bibinfo {year} {2021})}\BibitemShut {NoStop}%
\bibitem [{\citenamefont {Xia}\ and\ \citenamefont
  {Xu}(2015)}]{xia2015nonlinear}%
  \BibitemOpen
  \bibfield  {author} {\bibinfo {author} {\bibfnamefont {T.}~\bibnamefont
  {Xia}}\ and\ \bibinfo {author} {\bibfnamefont {X.}~\bibnamefont {Xu}},\
  }\href@noop {} {\bibfield  {journal} {\bibinfo  {journal} {Nuclear Fusion}\
  }\textbf {\bibinfo {volume} {55}},\ \bibinfo {pages} {113030} (\bibinfo
  {year} {2015})}\BibitemShut {NoStop}%
\bibitem [{\citenamefont {Rechester}\ and\ \citenamefont
  {Rosenbluth}(1978)}]{rechester1978electron}%
  \BibitemOpen
  \bibfield  {author} {\bibinfo {author} {\bibfnamefont {A.}~\bibnamefont
  {Rechester}}\ and\ \bibinfo {author} {\bibfnamefont {M.}~\bibnamefont
  {Rosenbluth}},\ }\href@noop {} {\bibfield  {journal} {\bibinfo  {journal}
  {Physical Review Letters}\ }\textbf {\bibinfo {volume} {40}},\ \bibinfo
  {pages} {38} (\bibinfo {year} {1978})}\BibitemShut {NoStop}%
\bibitem [{\citenamefont {Doerk}\ \emph {et~al.}(2012)\citenamefont {Doerk},
  \citenamefont {Jenko}, \citenamefont {G{\"o}rler}, \citenamefont {Told},
  \citenamefont {Pueschel},\ and\ \citenamefont
  {Hatch}}]{doerk2012gyrokinetic}%
  \BibitemOpen
  \bibfield  {author} {\bibinfo {author} {\bibfnamefont {H.}~\bibnamefont
  {Doerk}}, \bibinfo {author} {\bibfnamefont {F.}~\bibnamefont {Jenko}},
  \bibinfo {author} {\bibfnamefont {T.}~\bibnamefont {G{\"o}rler}}, \bibinfo
  {author} {\bibfnamefont {D.}~\bibnamefont {Told}}, \bibinfo {author}
  {\bibfnamefont {M.}~\bibnamefont {Pueschel}}, \ and\ \bibinfo {author}
  {\bibfnamefont {D.}~\bibnamefont {Hatch}},\ }\href@noop {} {\bibfield
  {journal} {\bibinfo  {journal} {Physics of Plasmas}\ }\textbf {\bibinfo
  {volume} {19}} (\bibinfo {year} {2012})}\BibitemShut {NoStop}%
\bibitem [{\citenamefont {Kaye}\ \emph {et~al.}(2007)\citenamefont {Kaye},
  \citenamefont {Levinton}, \citenamefont {Stutman}, \citenamefont {Tritz},
  \citenamefont {Yuh}, \citenamefont {Bell}, \citenamefont {Bell},
  \citenamefont {Domier}, \citenamefont {Gates}, \citenamefont {Horton} \emph
  {et~al.}}]{kaye2007confinement}%
  \BibitemOpen
  \bibfield  {author} {\bibinfo {author} {\bibfnamefont {S.}~\bibnamefont
  {Kaye}}, \bibinfo {author} {\bibfnamefont {F.}~\bibnamefont {Levinton}},
  \bibinfo {author} {\bibfnamefont {D.}~\bibnamefont {Stutman}}, \bibinfo
  {author} {\bibfnamefont {K.}~\bibnamefont {Tritz}}, \bibinfo {author}
  {\bibfnamefont {H.}~\bibnamefont {Yuh}}, \bibinfo {author} {\bibfnamefont
  {M.}~\bibnamefont {Bell}}, \bibinfo {author} {\bibfnamefont {R.}~\bibnamefont
  {Bell}}, \bibinfo {author} {\bibfnamefont {C.}~\bibnamefont {Domier}},
  \bibinfo {author} {\bibfnamefont {D.}~\bibnamefont {Gates}}, \bibinfo
  {author} {\bibfnamefont {W.}~\bibnamefont {Horton}},  \emph {et~al.},\
  }\href@noop {} {\bibfield  {journal} {\bibinfo  {journal} {Nuclear Fusion}\
  }\textbf {\bibinfo {volume} {47}},\ \bibinfo {pages} {499} (\bibinfo {year}
  {2007})}\BibitemShut {NoStop}%
\bibitem [{\citenamefont {Curie}\ \emph
  {et~al.}(2022{\natexlab{b}})\citenamefont {Curie}, \citenamefont {Hatch},
  \citenamefont {Halfmoon}, \citenamefont {Chen}, \citenamefont {Brower},
  \citenamefont {Hassan}, \citenamefont {Kotschenreuther}, \citenamefont
  {Mahajan}, \citenamefont {Groebner}, \citenamefont {team} \emph
  {et~al.}}]{curie2022gyrokinetic}%
  \BibitemOpen
  \bibfield  {author} {\bibinfo {author} {\bibfnamefont {M.}~\bibnamefont
  {Curie}}, \bibinfo {author} {\bibfnamefont {D.}~\bibnamefont {Hatch}},
  \bibinfo {author} {\bibfnamefont {M.}~\bibnamefont {Halfmoon}}, \bibinfo
  {author} {\bibfnamefont {J.}~\bibnamefont {Chen}}, \bibinfo {author}
  {\bibfnamefont {D.}~\bibnamefont {Brower}}, \bibinfo {author} {\bibfnamefont
  {E.}~\bibnamefont {Hassan}}, \bibinfo {author} {\bibfnamefont
  {M.}~\bibnamefont {Kotschenreuther}}, \bibinfo {author} {\bibfnamefont
  {S.}~\bibnamefont {Mahajan}}, \bibinfo {author} {\bibfnamefont
  {R.}~\bibnamefont {Groebner}}, \bibinfo {author} {\bibfnamefont {D.-D.}\
  \bibnamefont {team}},  \emph {et~al.},\ }\href@noop {} {\bibfield  {journal}
  {\bibinfo  {journal} {Nuclear Fusion}\ }\textbf {\bibinfo {volume} {62}},\
  \bibinfo {pages} {126061} (\bibinfo {year} {2022}{\natexlab{b}})}\BibitemShut
  {NoStop}%
\bibitem [{\citenamefont {Wang}\ \emph {et~al.}(2007)\citenamefont {Wang},
  \citenamefont {Hahm}, \citenamefont {Lee}, \citenamefont {Rewoldt},
  \citenamefont {Manickam},\ and\ \citenamefont {Tang}}]{wang2007nonlocal}%
  \BibitemOpen
  \bibfield  {author} {\bibinfo {author} {\bibfnamefont {W.}~\bibnamefont
  {Wang}}, \bibinfo {author} {\bibfnamefont {T.}~\bibnamefont {Hahm}}, \bibinfo
  {author} {\bibfnamefont {W.}~\bibnamefont {Lee}}, \bibinfo {author}
  {\bibfnamefont {G.}~\bibnamefont {Rewoldt}}, \bibinfo {author} {\bibfnamefont
  {J.}~\bibnamefont {Manickam}}, \ and\ \bibinfo {author} {\bibfnamefont
  {W.}~\bibnamefont {Tang}},\ }\href@noop {} {\bibfield  {journal} {\bibinfo
  {journal} {Physics of plasmas}\ }\textbf {\bibinfo {volume} {14}} (\bibinfo
  {year} {2007})}\BibitemShut {NoStop}%
\bibitem [{\citenamefont {Chen}\ \emph {et~al.}(2019)\citenamefont {Chen},
  \citenamefont {Xu},\ and\ \citenamefont {Lei}}]{chen2019extension}%
  \BibitemOpen
  \bibfield  {author} {\bibinfo {author} {\bibfnamefont {J.}~\bibnamefont
  {Chen}}, \bibinfo {author} {\bibfnamefont {X.}~\bibnamefont {Xu}}, \ and\
  \bibinfo {author} {\bibfnamefont {Y.}~\bibnamefont {Lei}},\ }\href@noop {}
  {\bibfield  {journal} {\bibinfo  {journal} {Computer Physics Communications}\
  }\textbf {\bibinfo {volume} {236}},\ \bibinfo {pages} {128} (\bibinfo {year}
  {2019})}\BibitemShut {NoStop}%
\bibitem [{\citenamefont {Hunana}\ \emph {et~al.}(2018)\citenamefont {Hunana},
  \citenamefont {Zank}, \citenamefont {Laurenza}, \citenamefont {Tenerani},
  \citenamefont {Webb}, \citenamefont {Goldstein}, \citenamefont {Velli},\ and\
  \citenamefont {Adhikari}}]{hunana2018new}%
  \BibitemOpen
  \bibfield  {author} {\bibinfo {author} {\bibfnamefont {P.}~\bibnamefont
  {Hunana}}, \bibinfo {author} {\bibfnamefont {G.}~\bibnamefont {Zank}},
  \bibinfo {author} {\bibfnamefont {M.}~\bibnamefont {Laurenza}}, \bibinfo
  {author} {\bibfnamefont {A.}~\bibnamefont {Tenerani}}, \bibinfo {author}
  {\bibfnamefont {G.}~\bibnamefont {Webb}}, \bibinfo {author} {\bibfnamefont
  {M.}~\bibnamefont {Goldstein}}, \bibinfo {author} {\bibfnamefont
  {M.}~\bibnamefont {Velli}}, \ and\ \bibinfo {author} {\bibfnamefont
  {L.}~\bibnamefont {Adhikari}},\ }\href@noop {} {\bibfield  {journal}
  {\bibinfo  {journal} {Physical review letters}\ }\textbf {\bibinfo {volume}
  {121}},\ \bibinfo {pages} {135101} (\bibinfo {year} {2018})}\BibitemShut
  {NoStop}%
\bibitem [{\citenamefont {Wang}\ \emph {et~al.}(2019)\citenamefont {Wang},
  \citenamefont {Zhu}, \citenamefont {Xu},\ and\ \citenamefont
  {Li}}]{wang2019landau}%
  \BibitemOpen
  \bibfield  {author} {\bibinfo {author} {\bibfnamefont {L.}~\bibnamefont
  {Wang}}, \bibinfo {author} {\bibfnamefont {B.}~\bibnamefont {Zhu}}, \bibinfo
  {author} {\bibfnamefont {X.-q.}\ \bibnamefont {Xu}}, \ and\ \bibinfo {author}
  {\bibfnamefont {B.}~\bibnamefont {Li}},\ }\href@noop {} {\bibfield  {journal}
  {\bibinfo  {journal} {AIP Advances}\ }\textbf {\bibinfo {volume} {9}}
  (\bibinfo {year} {2019})}\BibitemShut {NoStop}%
\bibitem [{\citenamefont {Wang}\ \emph {et~al.}(2020)\citenamefont {Wang},
  \citenamefont {Xu}, \citenamefont {Zhu}, \citenamefont {Ma},\ and\
  \citenamefont {Lei}}]{wang2020deep}%
  \BibitemOpen
  \bibfield  {author} {\bibinfo {author} {\bibfnamefont {L.}~\bibnamefont
  {Wang}}, \bibinfo {author} {\bibfnamefont {X.}~\bibnamefont {Xu}}, \bibinfo
  {author} {\bibfnamefont {B.}~\bibnamefont {Zhu}}, \bibinfo {author}
  {\bibfnamefont {C.}~\bibnamefont {Ma}}, \ and\ \bibinfo {author}
  {\bibfnamefont {Y.-a.}\ \bibnamefont {Lei}},\ }\href@noop {} {\bibfield
  {journal} {\bibinfo  {journal} {AIP Advances}\ }\textbf {\bibinfo {volume}
  {10}} (\bibinfo {year} {2020})}\BibitemShut {NoStop}%
\end{thebibliography}%

\end{document}